\documentclass[pdftex,twocolumn,epjc3]{svjour3}          

\RequirePackage[T1]{fontenc}

\smartqed  
\RequirePackage{multirow}
\RequirePackage{graphicx}
\RequirePackage{mathptmx}      
\RequirePackage{flushend}
\RequirePackage[numbers,sort&compress]{natbib}
\RequirePackage[colorlinks,citecolor=blue,urlcolor=blue,linkcolor=blue]{hyperref}

\usepackage{graphicx}  
\usepackage{dcolumn}   
\usepackage{bm}        
\usepackage{amssymb}   
\usepackage{feynmf}
\usepackage{hyperref}
\usepackage{slashed}
\usepackage{multirow}
\usepackage{color}
\usepackage{array}
\usepackage{caption}
\usepackage{subcaption}
\usepackage{amsmath}
\usepackage{booktabs}
\usepackage{multirow}
\usepackage{booktabs}
\usepackage{soul}

\newcommand{\oc}{OC}
\newcommand{\tspn}{TSPN-SA}

\begin{document}

\title{Reconstructing particles in jets using set transformer and hypergraph prediction networks}

\author{
    Francesco Armando Di Bello \thanksref{gen_add, e1}
    \and Etienne Dreyer \thanksref{wis_add, e2}
    \and Sanmay Ganguly \thanksref{tok_add}
    \and Eilam Gross \thanksref{wis_add}
    \and Lukas Heinrich \thanksref{tum_add}
    \and Anna Ivina \thanksref{wis_add}
    \and Marumi Kado \thanksref{mpi_add, rom_add}
    \and Nilotpal Kakati \thanksref{wis_add, e3}
    \and Lorenzo Santi \thanksref{rom_add}
    \and Jonathan Shlomi \thanksref{wis_add}
    \and Matteo Tusoni \thanksref{rom_add}
}

\thankstext{e1}{e-mail: francescoarmando.dibello@unige.it}
\thankstext{e2}{e-mail: etienne.dreyer@weizmann.ac.il}
\thankstext{e3}{e-mail: nilotpal.kakati@weizmann.ac.il}

\institute{
    INFN and University of Genova \label{gen_add}
    \and Weizmann Institute of Science \label{wis_add}
    \and ICEPP, University of Tokyo \label{tok_add}
    \and Technical University of Munich \label{tum_add}
    \and Max Planck Institute for Physics \label{mpi_add}
    \and INFN and Sapienza University of Rome  \label{rom_add}
}



\date{Received: date / Accepted: date}

\maketitle

\begin{abstract}

The task of reconstructing particles from low-level detector response data to predict the set of final state particles in collision events represents a set-to-set prediction task requiring the use of multiple features and 
their correlations in the input data. We deploy three separate set-to-set neural network architectures to reconstruct particles in events containing a single jet in a fully-simulated calorimeter. Performance is evaluated in terms of particle reconstruction quality, properties regression, and jet-level metrics. The results demonstrate that such a high-dimensional end-to-end approach succeeds in surpassing basic parametric approaches in disentangling individual neutral particles inside of jets and optimizing the use of complementary detector information.
In particular, the performance comparison favors a novel architecture based on learning hypergraph structure, \textit{HGPflow}, which benefits from a physically-interpretable approach to particle reconstruction.

\end{abstract}

\section{Introduction}


Testing theories in high energy physics rely on the ability to reconstruct high energy particle collision events using information recorded by particle detectors. General-purpose detectors enable this primarily through two sources of information: charged particle trajectories (tracks) measured in an inner tracking region and energy deposited by particle showers in a surrounding array of calorimeter cells.

Currently, experiments at the CERN Large Hadron Collider (LHC) employ parameterized ``particle-flow'' algorithms, which combine track and calorimeter information in a complementary way while avoiding double counting. 

The performance of particle-flow algorithms is limited to an extent by detector design specifications, such as the precision and size of the inner tracking system, the magnetic field strength in the tracking volume, the granularity of the calorimeters, and their energy resolution. 
However, a number of intrinsic factors complicate the task of particle reconstruction in the LHC environment: the busy and often collimated signatures resulting from proton collisions, the presence of multiple simultaneous scattering events (pileup), and finally, the extensive and irregular array of sensitive elements required for granularity and angular coverage.

\begin{figure*}[ht!]
    \centering
    \includegraphics[width=0.95\textwidth]{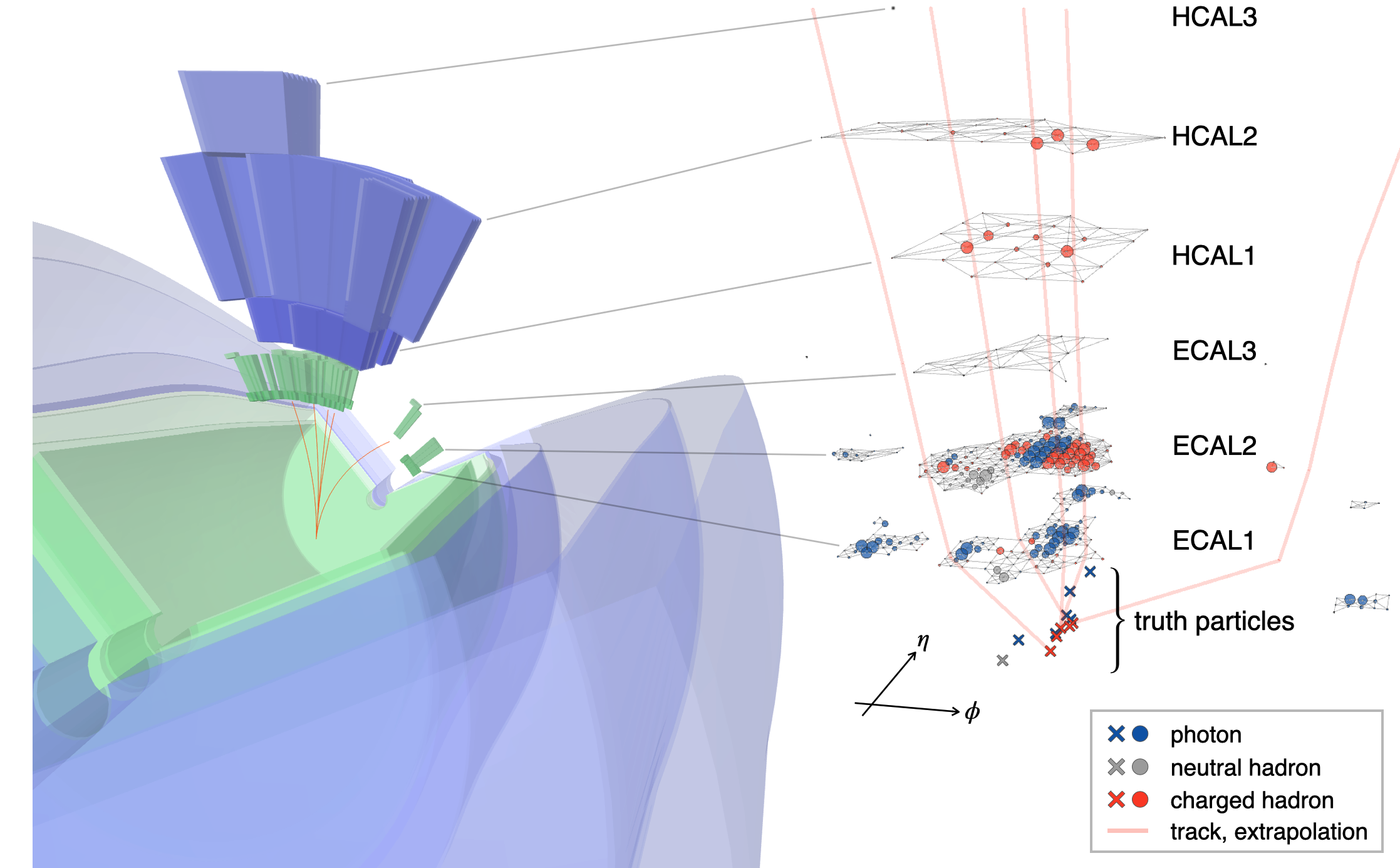}
    \caption{A depiction of a single-jet event from the test dataset in both the COCOA calorimeter layers (left) and as an input graph in $\eta-\phi$ space (right). On the left, the actual geometry of the calorimeter cells is shown, while on the right, they are represented by spheres with sizes proportional to their energy divided by noise threshold (up to a maximum value). Lines represent tracks and their projected locations in $\eta$ and $\phi$ in each calorimeter layer. Connections between calorimeter cells are the edges formed during graph construction (inter-layer edges and track-cell edges are not shown). The markers at the bottom right indicate the $\eta-\phi$ coordinates of the truth particles.}
\label{fig:graph_construct}
\end{figure*}

There are two main approaches to particle-flow algorithms. The approach used by the ATLAS collaboration~\cite{ATLAS:2017ghe} involves subtracting the expected shower profile for each track in an event from the calorimeter deposits to infer the energy contributed by nearby neutral particles. The CMS collaboration, on the other hand, employs a \textit{global} particle-flow algorithm where final state particles of different types are reconstructed simultaneously \cite{sirunyan2017particle}. Global particle-flow algorithms allow a high physics analysis flexibility and eliminate the need for overlap-removal algorithms while better exploiting the strengths of each sub-detector system. 

In this paper, we approach the global particle-flow paradigm using machine learning (ML) models operating on graph data. As in other applications to particle physics, ML brings the advantage of replacing parameterized cuts (for example, in energy subtraction schemes) with fully differentiable decision boundaries in the full space of relevant features in data. The expressiveness of ML models also opens new possibilities, such as reconstructing individual neutral particles inside of jets. Similarly, the choice to represent input data as graphs is motivated by several advantages: graphs more naturally capture the spatial correlations encoded in irregular detector geometry and also are well-suited for the sparsity and variable cardinality of the input set. Graph neural networks (GNN) have therefore emerged as an architecture of choice in recent particle reconstruction models, as they have in other particle physics tasks \cite{Shlomi_2020}.

In a collision event, the true set of particles $T$ upstream to the detector sensitive volume gives rise to a set of detector-level hits $D$. So the input set comprising the detector record is sampled from $p(D|T)$. Then global particle-flow reconstruction is the set-to-set task where the input set of detector-level hits $D$ is transformed into a typically much smaller output set $R$ comprising $N_R$ predicted particles. The predictions of a successful reconstruction algorithm $R(D)$ will correctly model the cardinality $N_T$ of $T$ and the properties (class, momentum, and angular coordinates) of its members. Several ML approaches have been proposed in the literature to predict $R(D)$.

In~\cite{Kieseler_2020} the object condensation (\oc{}) approach was proposed, which clusters nodes or pixels in latent space to form candidate objects, in our case, particles. Recently, \oc{} has been used to predict clusters in CMS data \cite{qasim2021multi,qasim2022end}, where the authors focused on reconstruction efficiency and energy regression of showers from single particles embedded in pileup. We implement \oc{} with modifications as explained in section \ref{sec:condensation} for the purpose of establishing a performance baseline for an ML-based particle reconstruction.

The reduction in size from input to output set is handled in the MLPF \cite{pata2021mlpf} approach by assigning input nodes to particle classes in the output set or else to a dedicated ``neglect class''. This approach was also recently successfully tested using CMS data~\cite{Pata:2022wam,Mokhtar:2023fzl}, where the model predictions were trained to match the output candidates from a standard particle-flow algorithm. For predicting true particles, MLPF is limited to cases where one or more clusters can be associated to each particle. It would therefore be required to define a fractional target definition in order to efficiently reconstruct particles that do not contribute a dominant fraction of energy in any single cluster (for example, a significant percentage of low-$p_\mathrm{T}$ photons).

In this paper, we contribute to the exploration of GNN-based particle reconstruction by proposing two new algorithms and comparing their performance alongside a modified \oc{} implementation as a baseline and a parameterized particle flow algorithm. Rather than full proton-proton collision events, we focus on events comprising a single jet, which represent the typical domain over which inter-particle correlations are expected to play a significant role in reconstruction. Our dataset incorporates full \textsc{Geant4}~\cite{geant} treatment of particle showers in a nearly-hermetic calorimeter simulation \cite{COCOA}. 
An example of a simulated single jet event is shown in Fig.~\ref{fig:graph_construct}. 
In the true particle-flow paradigm, our approach is built around the idea of combining low-level features from calorimeter showers with the complementary information provided by tracks. 

We find that a novel application of recurrent hypergraph learning leads to the most accurate results and preserves a high degree of interpretability. This is achieved thanks to a physics-inspired approach which allows the network to exploit the relationships between properties of the target particles and their energy deposits in the detector.



\section{Dataset}

\subsection{Detector simulation}
\label{sec:detector}

Unlike the full detector models used to simulate experiments at the LHC, publicly-available codes such as \textsc{Delphes} \cite{de2014delphes} do not model particle interactions with sufficient complexity to enable training a network with the full calorimeter signature available at real detectors. This motivated the development of the Configurable Calorimeter simulatiOn for AI (COCOA) package \cite{COCOA}, which we used to generate the datasets in this paper.


%
%
%
%
The geometric coverage of the COCOA calorimeter is split into a barrel ($0.0<|\eta|<1.5$) and two identical endcaps ($1.5<|\eta|<3.0$) regions. The endcap region is situated in a hermetic way such that there is no void in the transition region. In depth, the calorimeter has a total of six concentric layers: the first three layers comprising an electromagnetic calorimeter (ECAL) and the next three a hadronic calorimeter (HCAL). The calorimeters have uniform segmentation in $\eta$ and $\phi$ enabling high spatial resolution, as listed in Tab. \ref{tab:CaloLaeyrs}. The geometric depth of the cells is modulated as $1\,/\,\cosh{\eta}$ in order to achieve a constant effective interaction depth with increasing $\eta$.

The inner region of COCOA is immersed in a uniform axial magnetic field of 3.8T that extends until a radius of $150$ cm, where four $1.1$ cm layers of iron immediately precede the ECAL. The ECAL is modeled as a homogeneous calorimeter by mixing lead and liquid argon, corresponding to ATLAS calorimeter materials, in volume proportion $1.2:4.6$ leading to a radiation length of $X_0 = 2.5$ cm. For the HCAL, iron is used as the absorber material, and polyvinyl toluene plastic material as the scintillating active material. These are mixed with a volume proportion $1.1:1.0$, yielding a nuclear interaction length of $\lambda_{int}=26.6$ cm. The simulated energy deposits in each layer are smeared to reproduce the expected sampling energy resolution. For our dataset, the hadronic sampling term is 10\%. The effect of pileup and electronic noise is mimicked using normal distributions centered at zero with widths varying according to the layer.
The choice of material and smearing parameters is tuned to reproduce the ATLAS calorimeter system's single-particle response.

The effect of tracking is emulated by smearing truth charged particles with a resolution $\frac{\sigma(p)}{p} = a \times p $ with $a = 10^{-5}/$GeV. The smearing of the track direction is neglected as it is expected to have a subdominant effect in our problem of interest.

Charged particles produced from hadrons decaying-in-flight above a transverse radius $ R > $ 75 mm (250 mm) in the barrel (endcap) have no tracks associated to them. To focus on the reconstruction of particles as they appear at the calorimeter, the dataset simulates photon conversions only at the stage of the iron layer prior to the calorimeters, while material interactions within the tracker are emulated solely by the track $q/p$ smearing. 

\begin{table}[bt]
    \centering
    \caption{Characteristics of the six calorimeter layers: depths in radiation length ($X_0$) and nuclear interaction length ($\lambda_{int}$), granularity, and standard deviations of the simulated noise distributions.}
    \label{tab:CaloLaeyrs}
    \begin{tabular}{cccc}

    \toprule
            Layer & Depth                               &  Granularity ($\eta \times \phi$)                      & Noise {[}MeV{]} \\ 

    \midrule
        ECAL1 & $4X_0$                            &  $256 \times 256$ &    13               \\
        ECAL2 & $16X_0$                           &  $256 \times 256$ &   34               \\
        ECAL3 & $2X_0$                            &  $128 \times 128$ &  41               \\ 
    \midrule
        HCAL1 & $1.5\lambda_{int}$                &  $64 \times 64$ &  75               \\
        HCAL2 & $4.1\lambda_{int}$                &  $64 \times 64$ &  50               \\
        HCAL3 & $1.8\lambda_{int}$                &  $32 \times 32$ &  25               \\ 

    \bottomrule
    \end{tabular}
\end{table}

\subsection{Dataset generation}\label{sec:dataset}

Event generation, followed by parton shower and hadronization is performed with \textsc{Pythia8}~\cite{Pythia8} with a single initial state quark or gluon particle. The parton initial energy is sampled in the range 10 GeV $-$ 200 GeV, and angular coordinates are distributed uniformly in the range $\eta \in [-2.5,2.5]$, $\phi \in [-\pi, \pi]$. Final state particles are interfaced with \textsc{Geant4} to simulate their interaction with material, both showering in the calorimeter and scattering and e.g. photon conversions in the iron layer preceding it. Additional pileup collisions were not simulated. The targets of the machine learning algorithms are final state stable particles with transverse momentum above 1 GeV, which reach the calorimeter.

In each event, a standard clustering algorithm is used to group calorimeter cells into ``topoclusters'' based on their proximity and deposited energy, following the algorithm described in ~\cite{ATLAS:2016krp} with minor modifications. First, an energy over noise ratio of $(\frac{E}{\sigma}) > 4.6$ is used to identify cluster seeds. For each seed, a two-stage search is performed in its vicinity to group neighboring cells with nonzero energy. The first search collects neighboring cells with an energy-to-noise threshold ratio above 2, and the second search further extends the clusters with cells that have energy above 0. Finally, the algorithm applies a set of rules to merge topoclusters sharing seed cells and split topoclusters formed by particles in close proximity.

A record is kept of contributing particles and their energy contribution to each cell. Electronic noise is simulated in the calorimeter cells at realistic levels and dominates a fraction of the clustered cell. A small fraction of topoclusters, therefore, consist purely of cells where noise was the dominant contributor. One or more such topoclusters are present in 23\% of the training events.

\begin{figure}[!ht]
    \centering
    \includegraphics[width=0.48\textwidth]{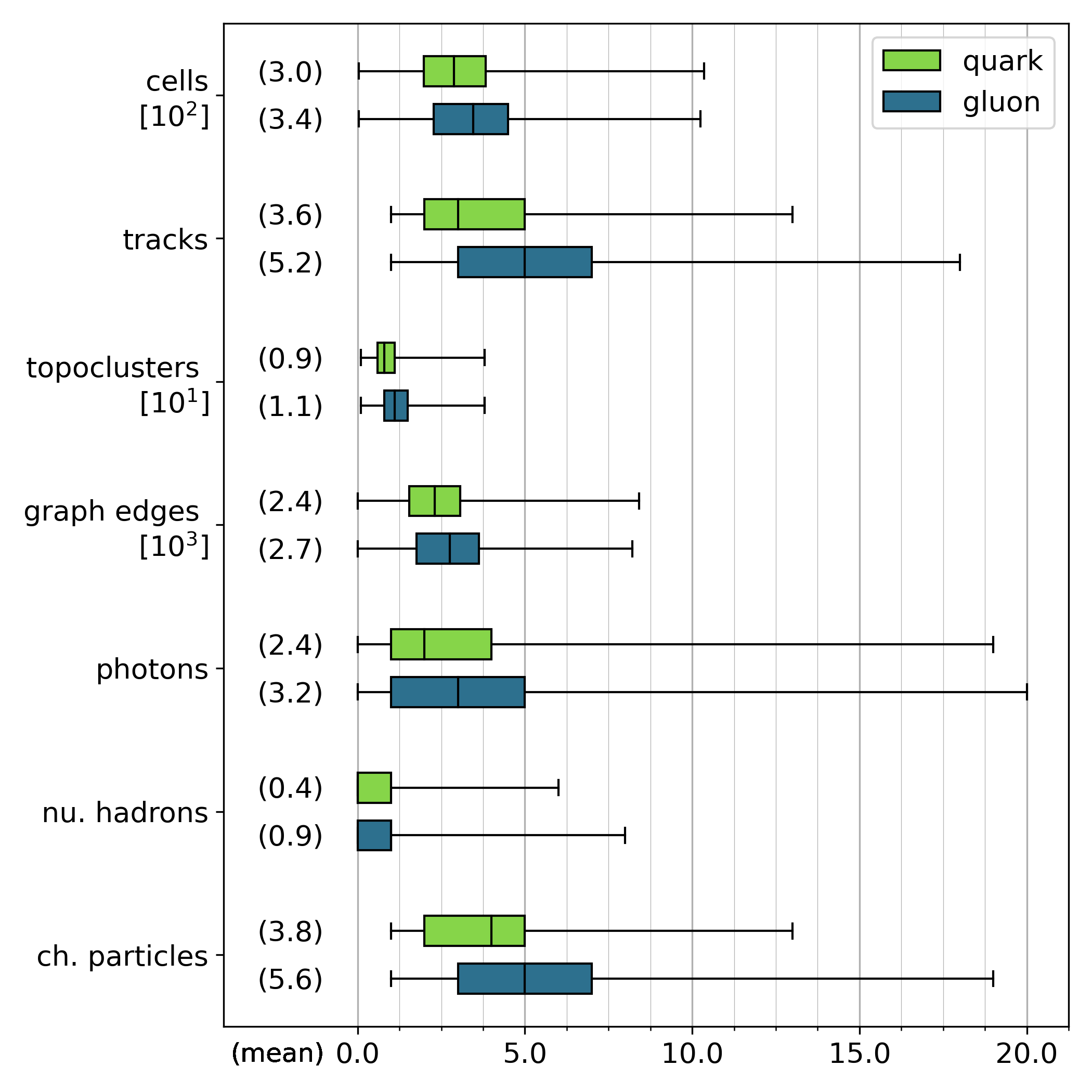}
    \caption{Composition of the quark and gluon test samples in terms of the cardinality of different sets of entities contained. The mean value in each case is written in parentheses while the range and quartiles of the distributions over events are shown in the box plot.}
    \label{fig:DataComposition}
\end{figure}

In summary, the data used for ML comprise the following object collections: cells which belong to a topocluster, all tracks that reached the calorimeter, and the set of particles which entered the calorimeter. An identical configuration is used to generate the dataset of 50000 events for training 
and the independent dataset of 30000 events for testing. In addition, a ``gluon jet'' dataset containing 30000 events is generated by replacing the single incident quark by a gluon with the same initial energy and angular distributions. The quark and gluon jet datasets are provided in \cite{note:zenodo}. The results obtained with this gluon jet sample are discussed in section \ref{sec:gluonperf}. Figure \ref{fig:DataComposition} summarizes the number of various entities stored in both the single jet and gluon jet test datasets.



\subsection{Fiducial particle definitions for reconstruction targets}
\label{sec:target}

In a collision event, not all particles produced can be reconstructed in the detector. When defining target truth particles it is important to account only for those that can be detected, {\it i.e.} those that are within the detector acceptance and have sufficiently high transverse momentum to be reconstructed. Beyond these simple criteria, particles produced in the collision can later decay or interact with the detector and convert, radiate or interact and produce other particles. The specific definition of the particles that are targets for the reconstruction algorithm, referred to as fiducial particles, is important to remove ambiguities during training and in assessing the performance of reconstruction. To qualify as fiducial particles, truth stable particles must have the following properties:
\begin{itemize}
    \item $p_\mathrm{T}$ > 1 GeV
    \item be produced before the first calorimeter layer
    \item release a nonzero amount of energy in the calorimeter
\end{itemize}


Additional consideration would be needed to achieve a more realistic environment where bremsstrahlung, pair production, and the presence of soft particles in general might result in highly collimated topologies, above the spatial reconstruction capabilities of the detector. In this work, the absence of pileup and the absence of material in the inner tracking region justify the use of the three fiducial criteria described above.


\begin{figure*}[!ht]
    \centering
    \includegraphics[width=0.75\textwidth]{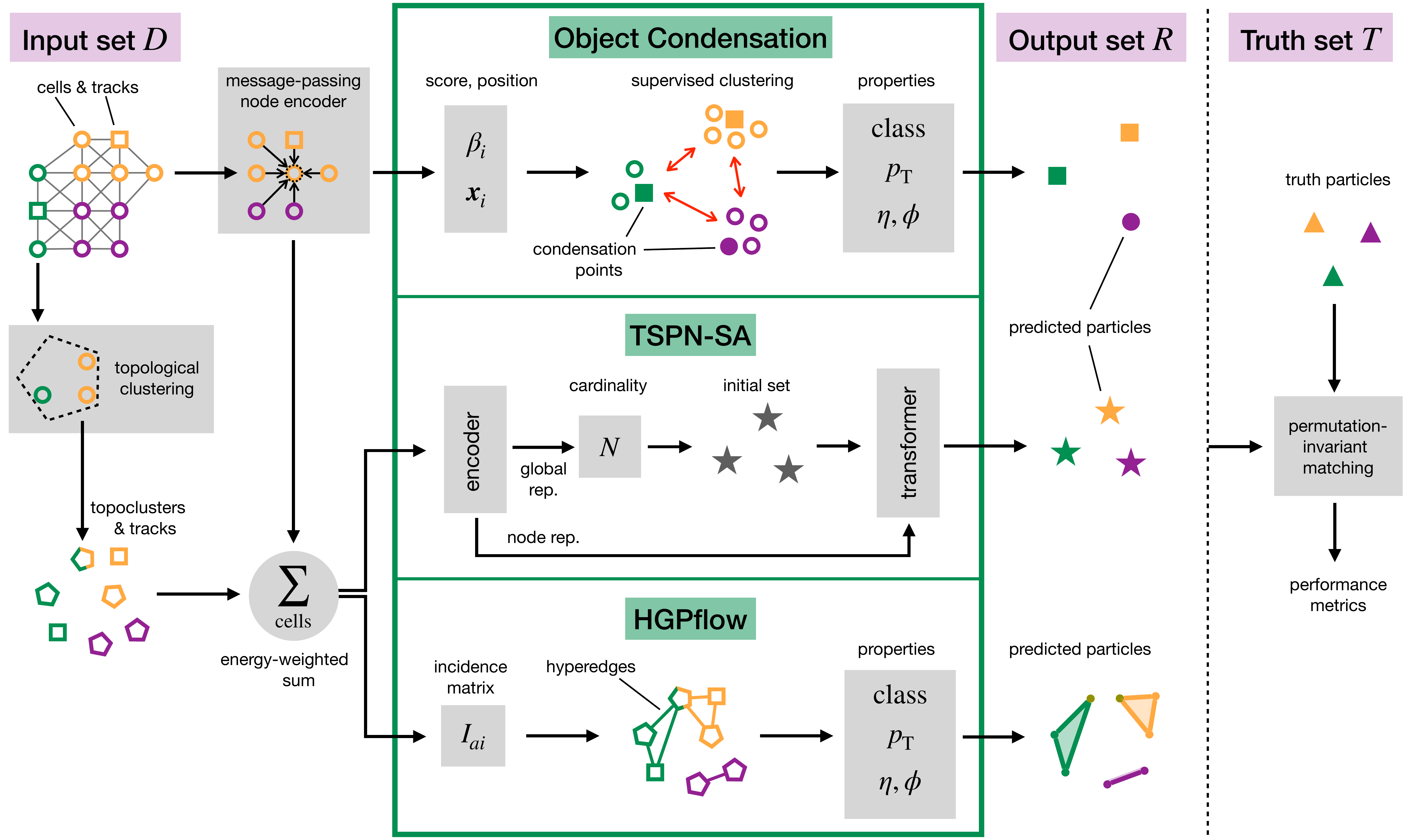}
    \caption{Comparison of the different ways in which the three ML reconstruction algorithms map the input set of nodes in the form of a graph to the output set of predicted particles, to be compared with the set of truth particles. Colors indicate distinct particles and the nodes for which they are the dominant contributor.}
    \label{fig:set2set}
\end{figure*}

\subsection{Input graph}\label{sec:graphconst}

We build a fixed heterogeneous graph out of each event by connecting calorimeter cells and tracks based on their proximity. Each cell is connected to the $k$ nearest cells in the same calorimeter layer, where ${k=8}$ in the ECAL and ${k=6}$ in the HCAL. Additionally, each cell is connected to the single nearest cell in its immediately adjacent layer(s). A cell in layer $l$ can only receive incoming edges from other cells if they are separated in $\Delta R$ by less than $\{d_\mathrm{max}^{c-c}\}_l = \{0.05, 0.07, 0.14, 0.30, 0.30, 0.60\}$ for the six calorimeter layers. A set of indices and weights is assigned per cell listing the true particles which contributed and their relative contribution to the total cell energy. An index of -1 is given to energy contributions from noise.

Tracks are likewise connected to cells based on closest separation in $\Delta R$ between the cell and the projected $\eta-\phi$ coordinate of the track in the corresponding calorimeter layer. A track is connected to a maximum number of ${k=4}$ cells in each ECAL layer and ${k=3}$ cells in each HCAL layer. For track-cell edges, a larger maximum $\Delta R$ is allowed: $\{d_\mathrm{max}^{t-c}\}_l = \{0.15,0.15,0.40,1.10,1.10,2.00\}$. A depiction of the graph connectivity for tracks and cells is shown in Fig.~\ref{fig:graph_construct}.

Topoclusters are represented in the input graph by a separate set of nodes with edges connecting each to the set of cells belonging to the topocluster. The angular coordinates of a topocluster are taken at its energy barycenter.

\section{Particle reconstruction algorithms} \label{sec:recoalgos}

\subsection{Parameterized particle-flow algorithm}\label{sec:ppflow}
To compare the performance of the ML algorithms, we implemented a traditional parameterised  particle-flow algorithm \cite{ATLAS:2017ghe}, which we refer to as PPflow. The algorithm aims at subtracting the energy deposited in the calorimeter from charged particles associated to tracks. To this end, shower templates are derived from single $\pi^+$ samples and parameterized as a function of the track $p_\mathrm{T}$ and the layer where the first nuclear interaction takes place. The energy subtraction is performed in concentric rings of radius equal to a single cell pitch built from the extrapolated track position in each calorimeter layer. The ring's energy is progressively subtracted from the topoclusters until the expected total energy determined in the single $\pi^+$ template is reached. The remaining energy in the topoclusters after this subtraction is considered as originating from neutral particles. The PPflow algorithm does not aim at reconstructing the single particles composing the jets, but rather it is designed to estimate the overall neutral energy component for each topocluster. 

\begin{figure*}[!ht]
    \centering
    \includegraphics[width=0.90\textwidth]{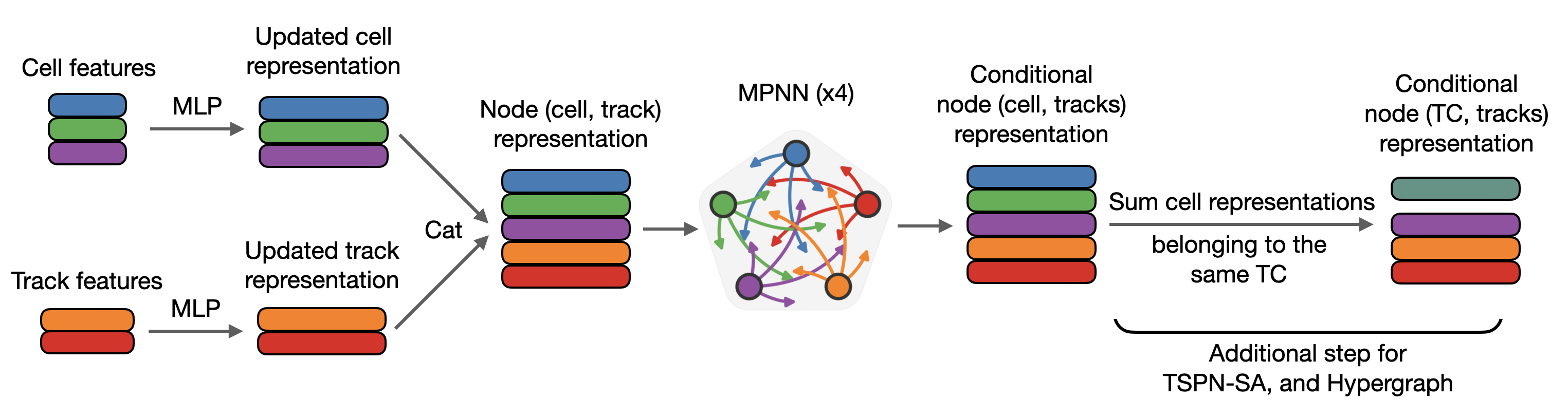}
    \caption{The encoding model used to derive a learned node representation.}
    \label{fig:model-encoding}
\end{figure*}

\subsection{Common description for the ML algorithms}\label{sec:models_common}




We investigate three ML-based particle reconstruction models for the set-to-set prediction $R(D)$: object condensation (\oc{}) as an existing ML baseline, transformer set prediction network with slot attention (\tspn{}), and a hypergraph architecture (HGPflow). Descriptions of each algorithm is given in sections \ref{sec:condensation}, \ref{sec:tspn}, and \ref{sec:hypergraph}. Section \ref{sec:performance} compares their relative particle-level performance and a comparison to the PPflow baseline for jet reconstruction.

There are commonalities to all three algorithms. Predicted particles in each case are inferred from the node features (skip connections) concatenated with a node representation vector from a common encoder network, discussed in section \ref{sec:emb}. Tracks are treated similarly in each case: the charged particle cardinality in an event is set by the number of tracks, and the charged particles' $\eta$ and $\phi$ are determined directly from the tracks without regression. While the \oc{} algorithm takes calorimeter cells as input nodes, the other two algorithms use topoclusters instead, to reduce the dimensionality of the input. Fig.~\ref{fig:set2set} illustrates the core differences between the ways each algorithm maps the set of detector-level nodes $D$ to the set of predicted particles $R$.

The choice to use calorimeter cells compared to using the coarser topoclusters can be compared in terms of an \textit{injective condition}: the degree to which the energy deposit in a node can be mapped back to a single parent particle. In the case of cells, although contributions from more than one parent particle are present in general, the injective condition is more valid than in the case of topoclusters. Since the injective condition is an assumption of the \oc{} algorithm (i.e. in the definition of the entries of $I_{ki}$ in Eq.~\ref{eq:L_V}), this motivates the choice of cells as input nodes.

Having contributions to a node from more than one parent particle can be learned in the \tspn{} architecture in an unsupervised way via node-particle attention. The HGPflow architecture, on the other hand, is fully equipped to disentangle multiple-particle contributions to a node thanks to supervised learning of the incidence matrix, discussed in section \ref{sec:hypergraph}. In both cases, computing gradients for predictions on edges becomes significantly more expensive for cell-level inputs compared to topocluster-level inputs, which was the main motivation for choosing the latter.

The loss associated with predicted particle properties is computed similarly in each algorithm. Particle class is trained using a categorical cross-entropy term between the predicted and the target class. A mean squared error loss term is used to regress continuous properties $\eta_i$ and $p_{\mathrm{T},i}$. The $\phi$ prediction is trained using $1 - \cos(\phi^\mathrm{pred} - \phi^\mathrm{targ})$.

The total number of trainable parameters in the neural network blocks of the \oc{}, \tspn{}, and HGPflow algorithms is compared in Tab. \ref{tab:ModelComparison} including the node encoding network in each. An estimate of their computational performance is also shown. For each of the three algorithms, hyperparameter optimization scans have not been performed, except on the threshold cuts used during inference for \oc{} and HGPflow. The code for the algorithms is provided in \cite{note:github}.

\begin{table}[bt]
    \caption{Comparison of the three particle reconstruction algorithms in terms of model size and computational resources. The number of trainable parameters belonging to the node encoding model is shown alongside the total. The time per event is averaged over 100 single jet events evaluated sequentially, and the memory is estimated as the peak memory over the same. Results are obtained on the same GPU (NVIDIA TITAN RTX).}
    \label{tab:ModelComparison}
    \begin{tabular}{c|l|c|c}

    \toprule
    \multirow{2}{*}{Algorithm} & \hspace{2.5mm} \# Parameters  & Speed* & Memory \\ 
      & Total \ (Node enc.) & [ms/event] & [MiB/event] \\
    \midrule
        \oc{}   & 1.8M \ (0.2M) & 249  & 1480\\
        \tspn{} & 1.5M \ (0.2M) & 465  & 1448\\
        HGPflow & 1.8M \ (0.2M) & 257  & 1394\\
    \bottomrule
    \end{tabular}
     \begin{flushright}\vspace{-1mm}*algorithms not optimized for execution time.\end{flushright}
\end{table}

%
%

\subsection{Graph nodes encoding}
\label{sec:emb}

Each event is represented as a heterogeneous graph comprising track, cell, and topocluster nodes connected by edges as defined in section \ref{sec:graphconst}. The embedding model described in the following is shared among the different network architectures. Fig.~\ref{fig:model-encoding} illustrates the network components of the encoding model: input feature vectors associated with track and cell nodes are passed through separate networks to embed them in a common representation space of dimension 100. The cell features input to the embedding are (energy, position, $\phi$, $\eta$, layer). Similarly, the track input features are the track parameters ($q/p$, $\theta$, $\phi$, $d_0$, $z_0$)\footnote{The track impact parameters $d_0$ and $z_0$ measure the distance of closest approach of the track to the beam line in the transverse and longitudinal directions, respectively.} and the extrapolated $\eta$-$\phi$ coordinates of the track at each calorimeter layer. The latter are important features because a charged particle after exiting the magnetic field travels in a straight line which no longer points back to the origin (assuming no material interactions). In addition to their hidden representation, these nodes are also given an additional binary feature which flags whether they originate from cells or tracks.

The node encodings are then updated to incorporate the graph relational structure via 4 successive blocks of message passing along edges. In each block, a dedicated network is used with the following three inputs concatenated: current node representation, sum of representations from neighboring nodes, and a graph-level global representation (the mean of all current node representations). Following the message passing blocks, topocluster representations are computed by the energy-weighted mean of the cell representation vectors belonging to the topocluster.


\begin{figure*}[!h]
     \centering
    \includegraphics[width=0.95\textwidth]{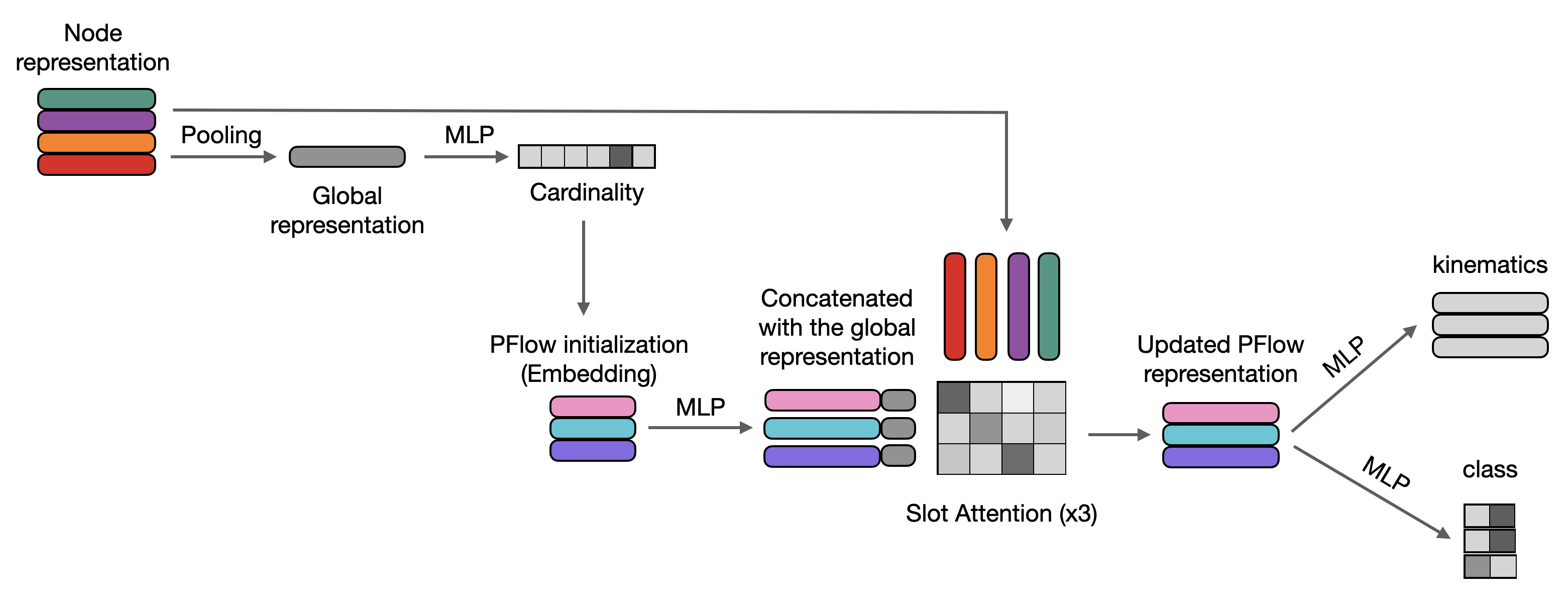}
\caption{The \tspn{} architecture. The cardinality of the set of output particles is predicted from the global representation, while their properties are predicted from representation vectors resulting from successive slot attention blocks.}
\label{fig:tspn_superneutral_all}
\end{figure*}

\subsection{Modified object condensation (\oc{})}\label{sec:condensation}

The \oc{} algorithm was proposed in~\cite{Kieseler_2020} for tasks of segmenting a set of input nodes into a set of target objects and prediction of their properties, which it does simultaneously. In the particle reconstruction case, the input set comprises tracks and calorimeter cells and the output set of objects are the progenitor particles (``parents'') with their classes and properties. The set-to-set procedure for \oc{} is shown in the top row of Fig.~\ref{fig:set2set}. Our implementation follows the original \oc{} approach with some modifications which are stated in the following  description. 

The \oc{} algorithm is based on clustering nodes according to their parents in a learned few-dimensional space $x$. The clustering is supervised by adding to the loss potentials defined on this space: a repulsive potential $\hat{V}(x) \propto \max(2-\Delta x,0)$ between nodes that belong to different parent particles, and an attractive one $\breve{V}(x) \propto \Delta x^2$ for nodes having a common parent. The goal is that after training the resulting clusters of nodes will correspond to the set of parent particles.

However, calculating the sum of $N^2$ pairwise potential terms during training becomes expensive for problems of even moderate $N$. This is addressed by designating a single representative node for each parent particle, called a \textit{condensation point}, to impose the potentials on all other nodes during training. A separate network is trained to predict a score, $\beta \in [0,1] $, with a target value of 1 for condensation points and 0 otherwise. An increasing function $q(\beta) = \mathrm{arctanh}^2 \beta + q_\mathrm{min}$ (with $q_\mathrm{min}$ a hyperparameter) is used in analogy to charge in the loss term responsible for the clustering potentials:
\begin{equation} \label{eq:L_V}
    L_V = \frac{1}{N} \sum_{i=1}^{N}q(\beta_i) \sum_{k=1}^{K}\left[I_{ki} \breve{V}_{c,k}(x_i) + (1-I_{ki})\hat{V}_{c,k}(x_i)\right]
\end{equation}
where $I_{ki}$ is an $N\times K$ matrix determining whether particle $k$ is the parent of node $i$. The matrix will be revisited in section \ref{sec:hypergraph}.

For each node $i$, which is a cell, the properties loss is of the same form as discussed in section \ref{sec:models_common}, where target class is either photon, neutral hadron or charged particle. Similar to $L_V$ above, the particle property loss is also weighted by $q(\beta)$ such that nodes with the highest $\beta$ receive the most supervision during training. These nodes are ultimately selected for the output set during inference by requiring their predicted $\beta > t_\beta$ and that they be separated in the clustering space by $\Delta x > t_d$, where $t_\beta$ and $t_d$ are two threshold hyperparameters.

Compared to the original \oc{} model, our implementation has two modifications connected to the condensation score $\beta$. The condensation points defined during training do not have a physical meaning and are learned in an unsupervised way. In our approach, we instead use the following physics-oriented definition:
\begin{equation}
\mathrm{CP}_{k}^{\ T} = 
    \begin{cases}
        \phantom{\mathrm{argmax}_{z})}      \mathrm{track} \in k ,\ \  \mathrm{if}\ k\  \mathrm{is\ charged\ particle} \\
        \mathrm{argmax}_{z}(\mathrm{cells} \in k),\  \mathrm{if}\  k\ \mathrm{is\ neutral\ particle}
    \end{cases} \label{eq:CPdef_train}
\end{equation}
where $z$ is the energy over noise threshold ratio for each cell. This definition removes the need to identify a representative node for charged particles, assuming a 1-1 mapping to tracks in the event. For neutral particles, on the other hand, the $\beta$ prediction is fully supervised and can be interpreted as the likelihood that a cell has maximal $z$ in a given shower. Since this cell also serves as an approximate location of the shower center, the $\eta$ and $\phi$ for neutral particles are regressed via a learned offset to the cell $\eta$-$\phi$ coordinates. During inference, condensation points passing the $t_b$ and $t_d$ thresholds are further required to be classified as either photon or neutral hadron, whereas cells classified as charged particles are discarded (since this role is fulfilled by tracks).

A second modification compared to the original \oc{} approach is that instead of a $\sim(1-\beta)$ regression-type loss computed on condensation points only, we train the $\beta$ prediction using a binary cross-entropy loss evaluated for all nodes. The reasoning behind a classification-type loss is to directly penalize the network for predicting large $\beta$ for nodes which are not condensation points, i.e. false positives. In an ablation study, each modification was seen to bring substantial improvement at essentially no additional model complexity.

Besides the two modifications above, our OC implementation differs from that of \cite{qasim2022end} in a few regards. Firstly, they propose an upgrade to the original \oc{} algorithm where particles are represented in the clustering space not only by a singular condensation point but by the $\beta$-weighted average over a learned distance scale. The authors of \cite{qasim2022end} report that these and other modifications lead to improved training stability and reduced noise, so we recognize the potential for improving our implementation as well. Finally, we point out that our model has significantly more parameters (Tab. \ref{tab:ModelComparison}), with GravNet blocks \cite{Qasim:2019otl} replaced by the node encoding model described in section \ref{sec:emb}. Our choice of network block sizes has not been optimized for computational efficiency and allocates a large proportion of its parameters to the node prediction networks compared to the message-passing networks.



\subsection{Transformer Set Prediction Network with Slot Attention (\tspn{})}\label{sec:tspn}

The Transformer Set Prediction Network (TSPN) was initially developed for the permutation-invariant encoding and decoding of variable-size sets of feature vectors~\cite{kosiorek2020conditional}. The utility of this for set-to-set problems in particle physics is clear: a model is needed that predicts an output set of entities (i.e. particles) based on an input set of different entities (i.e. calorimeter clusters, tracks), where both sets typically have different cardinality.
The model is divided into two networks: the first for predicting neutral particles and the second for predicting charged particles (discussed later).

As shown in the top part of Fig.~\ref{fig:tspn_superneutral_all}, the first architecture starts with a set encoder network whose output is used to predict the number of neutral particles, $N^\mathrm{pred}$
This prediction is trained using a categorical cross-entropy loss over 25 classes, which is an upper bound on the number of neutral particles per event. The cardinality prediction is used during test time. 
During training, the truth cardinality is enumerated to form a set of numbers that are passed through embedding layers to instantiate the initial set of random vectors. These vectors combined with the global representation vector function as queries for a series of 3 slot-attention (SA) blocks \cite{locatello2020object}. The SA blocks are not part of the original TSPN proposal but were found to be fundamental for performance in our application (abbreviated \tspn{}).
Each block contains 3 iterations where the particle hidden representation is updated using the attention-weighted representations of the topoclusters and tracks in the event.
Finally, the updated hidden representations are inputs to two dedicated neural network blocks aimed at predicting the kinematics ($p_\mathrm{T}$,$\eta$,$\phi$) and the class of the particle candidates.

The neutral particle properties loss function for the \tspn{} algorithm is defined as the sum of  a categorical cross-entropy term for class $t_i$  and a mean-squared error ($\mathrm{MSE}$) term for the continuous properties $\bm{p}_i=\{\eta_i,\phi_i,p_{\mathrm{T},i}\}$ computed for each particle candidate. 
The target particles are defined by matching to the set of predicted particles using the Hungarian assignment algorithm \cite{kuhn1955hungarian}, with the loss itself being the distance metric. 

 \begin{figure*}[ht!]
    \centering
    \includegraphics[width=0.95\textwidth]{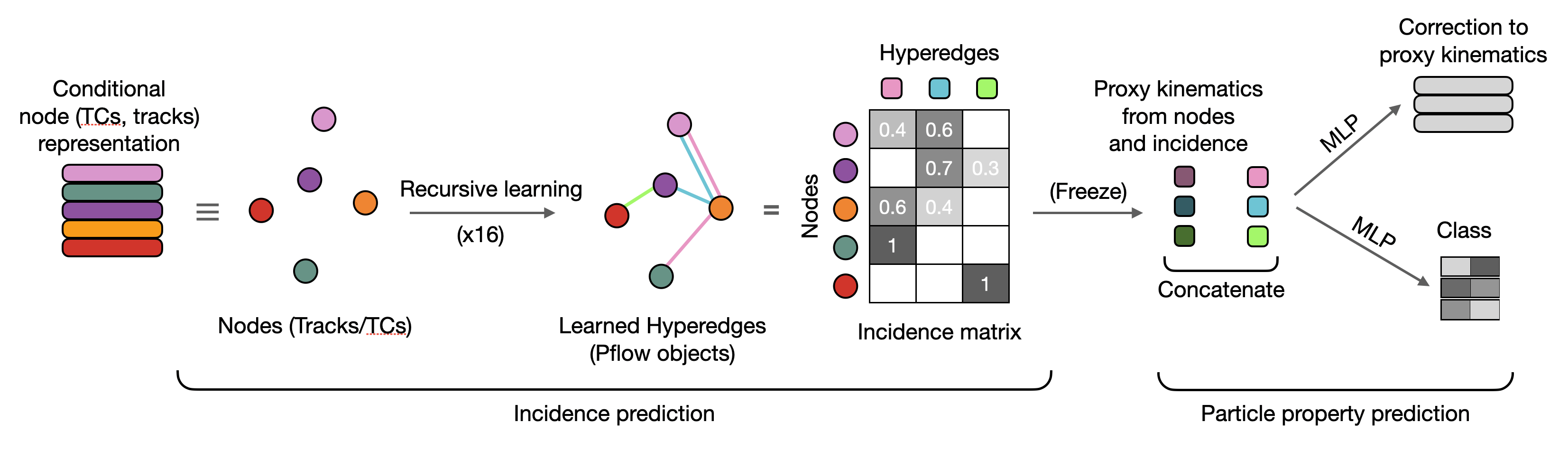}
    \caption{The two stages of learning in the HGPflow algorithm. The objective of the first stage is to predict the fractional entries in the energy-weighted incidence matrix, where columns correspond to hyperedges (i.e. particles). This is done by accumulating the loss over a sequence of recursive updates. In the second stage, the incidence matrix is frozen and the network minimizes losses for particle property predictions, defined relative to proxy quantities.}
    \label{fig:hypergraph}
\end{figure*}

The second part of the \tspn{} algorithm, for predicting charged particles, makes explicit use of the prior knowledge originating from track-particle objects. Each track is promoted to a particle, such that the cardinality of the output set is fixed by the number of tracks. Similar to the neutral architecture, SA blocks are used to update the hidden representation of the charged particle candidates. Unlike the model for neutral particles, the only predicted quantity for charged particles is their transverse momentum in order to improve over the track resolution. For charged particles, the track index is used to match with the corresponding target particle. As in the \oc{} algorithm, the $\eta$ and $\phi$ of the charged particles are taken directly from their representative tracks. Finally, the total loss for the \tspn{} algorithm is the sum of neutral particle cardinality loss, the neutral particle properties loss, and the charged particle $p_\mathrm{T}$ MSE loss. These loss terms are minimized simultaneously during training.



\subsection{HGPflow: particles as hyperedges}\label{sec:hypergraph}

\begin{figure}
\centering
    \includegraphics[width=0.5\textwidth]{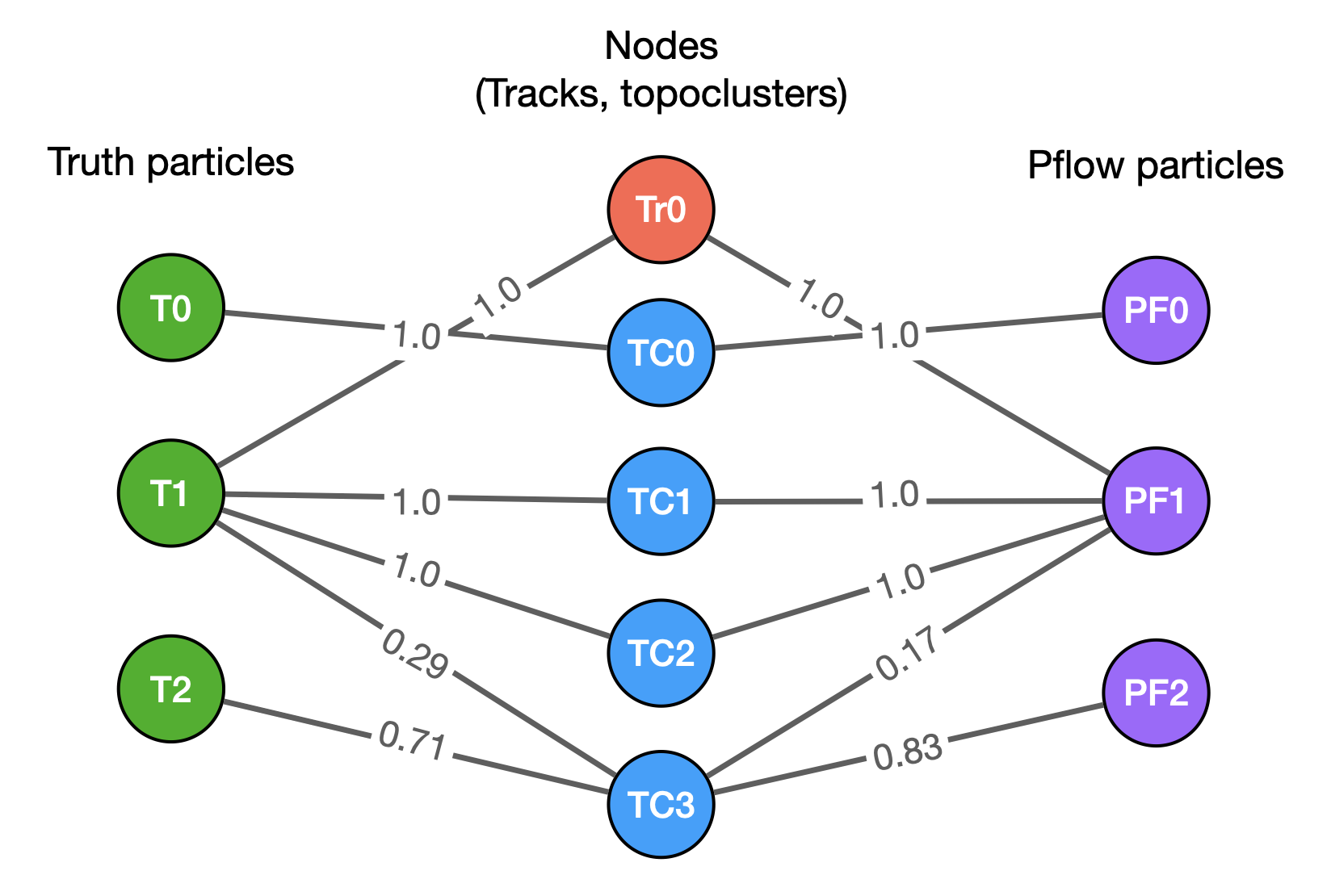}
\caption{Schematic representation of the truth and predicted incidence matrix in HGPflow for one event. The left part of the diagram shows the three truth particles in the event. One of them has a track (Tr) associated to it. The three particles deposit their energy into four topoclusters (TC). The links represent the fractional energy originated by a given particle in a given topocluster or track.  The right part of the diagram shows the predicted values of the incidence matrix for each reconstructed particle.}
\label{fig:incidence}
\end{figure}

A hypergraph is a generalization of a graph where hyperedges can each connect one, two, or multiple nodes (Fig.~\ref{fig:hypergraph}). 
While connectivity in a graph of $N$ nodes is described by an $N\times N$ adjacency matrix, a hypergraph containing $K$ hyperedges is described by an incidence matrix $I^{(N\times K)}$.
In the context of particle reconstruction, calorimeter deposits and tracks can be represented as nodes in a hypergraph, while each particle is represented by a hyperedge connecting the set of nodes to which it contributed. We propose \textit{HGPflow}: an algorithm that treats particle reconstruction as a task of learning hyperedges and their properties. There are two objectives in the training of HGPflow:
\begin{enumerate}
    \item predict the incidence matrix defining the hyperedges
    \item predict the hyperedge (i.e. particle) properties
\end{enumerate}
The first objective is similar to the task of separating overlapping charged and neutral showers which was the focus of ~\cite{DiBello:2020bas}. In this first stage, the HGPflow network predicts $(N+1) \times K$ entries comprising a zero-padded incidence matrix and an additional row of binary values that indicates whether the particle corresponding to a given column exists or not. Since the number of particles per event varies, the number of columns $K$ is set to an upper bound on the number of particles estimated from the training set (in our case $K=30$).
To express a non-injective map from particles to nodes, we define a target incidence matrix which has fractional rather than binary-valued entries. The entry relating node $i$ to particle $a$ is the following:
\begin{equation} \label{eq:incidence}
    [I]_{ia} = \frac{E_{ia}}{\sum\limits_{\mathrm{particles}\ b}E_{ib}} = \frac{E_{ia}}{E_i}
\end{equation}
where $E_{ia}$ is the amount of energy that particle $a$ contributes to the total energy $E_i$ of node $i$. For nodes which are tracks, incidence entries are simply 1, whereas for topoclusters they compute the fraction of the topocluster's energy that came from a given particle. An example of target and predicted incidence matrix entries are shown in Fig.~\ref{fig:incidence} for one event.

Predicted rows in the incidence matrix are normalized using Softmax (i.e. sum over all hyperedges for a given node is 1) before being compared to the target via Kullback–Leibler divergence loss.
\begin{equation}
    L_{\mathrm{inc}} = \sum_a \mathrm{KL}_i\left( I^\mathrm{targ}_{ia} , \mathrm{Softmax}_i(I^\mathrm{pred}_{ia}) \right) \label{eq:incidence_loss}
\end{equation}
The predicted entries of the indicator row are passed through a sigmoid function and compared to the (binary) target entries using a binary cross entropy loss function. Predicted columns are rearranged using the Hungarian algorithm to minimize the loss.

The incidence matrix prediction network is trained using the recurrent strategy proposed by \cite{zhang2021recurrently}, described briefly hereafter. The loss in Eq.~\ref{eq:incidence_loss} is calculated for a sequence of 16 refinement blocks each comprising an updated prediction of the incidence matrix followed by an update of node representation vectors $V$, and hyperedge representation vectors $E$. The iteration $t \rightarrow t+1$ is performed with the following three successive steps:
\begin{align}
I^{t+1}_{ia} &= \phi_I\left( v_i^t, e_a^t, I_{ia}^t \right) \label{eq:incidence_update} \\
V^{t+1} &= \phi_V\left( \left\{ v_i^t, \rho_{E\rightarrow V}(i,t) , v^0  \big| i=1\dots n \right\}\right) \label{eq:node_update} \\
E^{t+1} &= \phi_E\left( \left\{ e_a^t, \rho_{V\rightarrow E}(a,t) \hspace{3mm} \big| a=1\dots k \right\} \right) \label{eq:hyperedge_update}
\end{align}
where $\rho_{E\rightarrow V}(i,t) = \sum_{a} I_{ia}^{t+1} e_a^t$ and $ \rho_{V\rightarrow E}(a,t) = \sum_{i} I_{ia}^{t+1} v_i^t$ are aggregations of node ($v$) and hyperedge ($e$) representation vectors weighted by the updated incidence matrix. The updates are performed at each step using the same networks $\phi_I$, $\phi_V$, and $\phi_E$, where the latter two networks are DeepSets models \cite{zaheer2017deep}.

To reduce computational cost, not every iteration of the backward pass is included in the gradient step. Two sequences of 4 adjacent iterations are randomly selected out of the 16 for which the incidence loss is computed and added to the loss from the prediction at the end of the sequence.

The second training objective of the HGPflow network (Fig.~\ref{fig:hypergraph}c) is to predict particle properties for each hyperedge. 
The corresponding loss function contains classification and regression terms evaluated by matching predicted and target particles using the Hungarian algorithm. Particles corresponding to hyperedges where the predicted indicator was below the threshold are matched to dummy targets and weighted by zero in the loss.
Classification between photons and neutral hadrons is performed for hyperedges which do not contain a track and are thus identified as neutral particles.
The regression task benefits from a unique advantage enabled by learning the incidence matrix (Eq.~\ref{eq:incidence}): particle kinematics can be approximated as weighted sums and averages over the input features of the topoclusters contained in the hyperedge. Proxy quantities (denoted $\hat{}\ $) for energy and angular coordinates can be computed as:
\begin{equation} \label{eq:proxy}
\hat{E}_a = \sum\limits_{\mathrm{nodes}\ i} E_{i} I_{ia}\ , \hspace{5mm} \{\hat{\eta_a},\hat{\phi_a}\} =  \sum\limits_{\mathrm{nodes}\ i} { \{\eta_i,\phi_i\} \tilde{I}_{ia}}
\end{equation}
where a dual incidence matrix $\tilde{I}$, normalized over node instead of particle indices, can be defined:
\begin{equation}
    \tilde{I}_{ia} = \frac{E_{ia}}{\sum\limits_{\mathrm{nodes}\ j} E_{ja}} = \frac{E_{ia}}{E_a} = \frac{E_{i}\cdot I_{ia}}{\sum\limits_{\mathrm{nodes}\ j}  \left(E_{j}\cdot I_{ja}\right)}  \label{eq:dual}
\end{equation}
The property prediction networks in HGPflow are therefore given the simpler objective of learning corrections to the approximate values from Eq.~\ref{eq:proxy}. The loss terms used for the property predictions follow the description in section \ref{sec:models_common}.

Therefore, neutral particle kinematics ($p_\mathrm{T}, \eta, \phi$) are regressed by predicting an offset to the proxy values in Eq.~\ref{eq:proxy}. For charged particles, an offset is likewise predicted for the $p_\mathrm{T}$ measured from the associated track. The properties loss is computed by matching predicted and target particles using the Hungarian algorithm \cite{kuhn1955hungarian}.


\section{Performance of particle reconstruction in jets}\label{sec:performance}

One of the most challenging tasks of global particle flow algorithms is the reconstruction of particles in dense environments, in particular jets. In this section, the performance of the ML reconstruction algorithms will be assessed by quantifying the similarity between the set of predicted and set of target particles. The following four types of metrics are meant to evaluate the cardinality, class, and properties predictions:
\begin{itemize}
    \item Efficiency and fake rate
    \item Classification purity
    \item Particle angular and momentum resolution
    \item Jet-level quantities
\end{itemize}
The efficiency and fake rate are defined as follows:
\begin{equation}
    \epsilon \equiv \frac{N(\mathrm{matched\ pred})}{N(\mathrm{targ})} \hspace{2mm} ,
    \ f \equiv \frac{N(\mathrm{unmatched\ pred})}{N(\mathrm{pred})}
\end{equation}

The quality of the regression tasks is evaluated from distributions of their residuals, defined as $(y_\mathrm{targ} - y_\mathrm{pred})/y_\mathrm{targ}$ for particle property $y \in \{ p_\mathrm{T}, \eta, \phi \}$.

\begin{figure*}[ht!]
    \centering
    \includegraphics[width=0.6\textwidth]{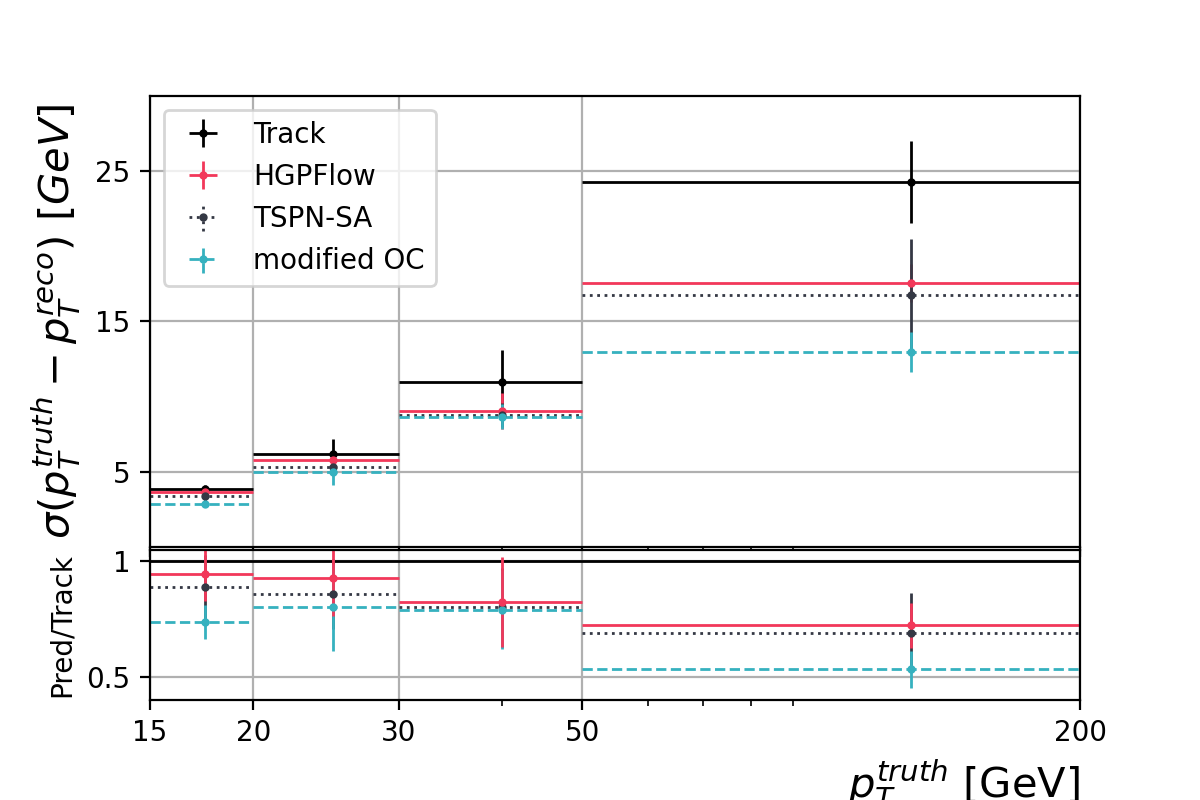}

\caption{Resolution of charged particle-flow candidates and tracks as a function of the associated particle transverse momentum.  At high $p_\mathrm{T}$  the particle-flow candidates show improved resolutions over the tracks.}
\label{fig:ch_reso}
\end{figure*}

\subsection{Particle matching}

Predicted and target particles are matched using the Hungarian algorithm to find the pairings which minimize the distance between their properties, defined by the following metric:
\begin{equation}
  d_\mathrm{match} = \sqrt{(\Delta p_\mathrm{T} / p_\mathrm{T}^\mathrm{truth})^2 + \Delta R^2}
\end{equation}
where $\Delta$ denotes the difference between a predicted and target property, and $\Delta R^2 = \Delta \eta^2 + \Delta \phi^2$. Matching is performed separately for neutral particles and charged particles since the latter are distinguished by the presence of a track. The coefficients $c_{p_\mathrm{T}}$ and $c_{\Delta R}$ are set to $1$ and $5$ for neutral particles while for charged, matching is based only on $\Delta R$ (i.e. $c_{p_\mathrm{T}}=0$). Prioritizing spatial matching helps decouple reconstruction efficiency from classification accuracy, which in particular will dominate at low-$p_\mathrm{T}$ because of the similarity of photon and neutral hadron signatures.

In each event, when the cardinality of the predicted set of particles is larger than that of the target, the non-matched predictions are labeled as ``fake'' particles. Conversely, inefficiency arises when not enough neutral or charged particles were predicted in order to match every target.



\subsection{Charged particle performance}

Charged particles include electrons, muons, and charged hadrons. In jets charged pions produced during hadronization account for around 90\% of all charged particles.  Leptons such as electrons and muons are present in less than 3\% of the jets. Electrons are produced from photon conversions and hadrons decaying in flight while muons are mostly produced by the latter mechanism.  Given the large class imbalance and the fact that no dedicated studies have been performed to improve the classification of electrons inside jets,  the three classes are grouped together and characterized as a single class of charged particles. 

Tracking efficiency presents an upper bound on the efficiency of charged particle reconstruction (see Section~\ref{sec:detector}). Since fake tracks are not emulated in the data, charged particle fake rates are neglected in this study. In any case, the rate of fake tracks at 1 GeV is typically at the percent level for the ATLAS and CMS experiments, which is expected to have a small impact.

In cases where the track belonging to a charged particle is not reconstructed, the target particle is relabelled to avoid confusing the network during training. Charged hadrons without a track are relabelled as neutral hadrons, and electrons without a track as photons. Photon pairs from neutral pion decays prior to the calorimeter are treated as two distinct target particles.

A key characteristic of charged particle reconstruction is the resolution of $p_\mathrm{T}$ with respect to the true value. It is well known that at low transverse momentum tracks provide the best momentum estimate over the calorimeter resolutions. An opposite trend appears at high energies where the calorimeter systems provide the most accurate energy measurement. 
Fig.~\ref{fig:ch_reso} shows the resolutions of charged particles reconstructed with the three ML approaches and compared to the track resolutions for charged particles with $p_\mathrm{T}$ > 15 GeV. Below this value, the particle $p_\mathrm{T}$ regression is replaced with simply the track $p_\mathrm{T}$ since an improvement is not expected. An increasing improvement at high $p_\mathrm{T}$ is observed for all the ML algorithms demonstrating that indeed the complementarity between the calorimeter and tracking measurements has been learned during training. 



\begin{figure*}[ht!]
\centering
\begin{subfigure}[b]{0.4\textwidth}
    \includegraphics[width=\textwidth]{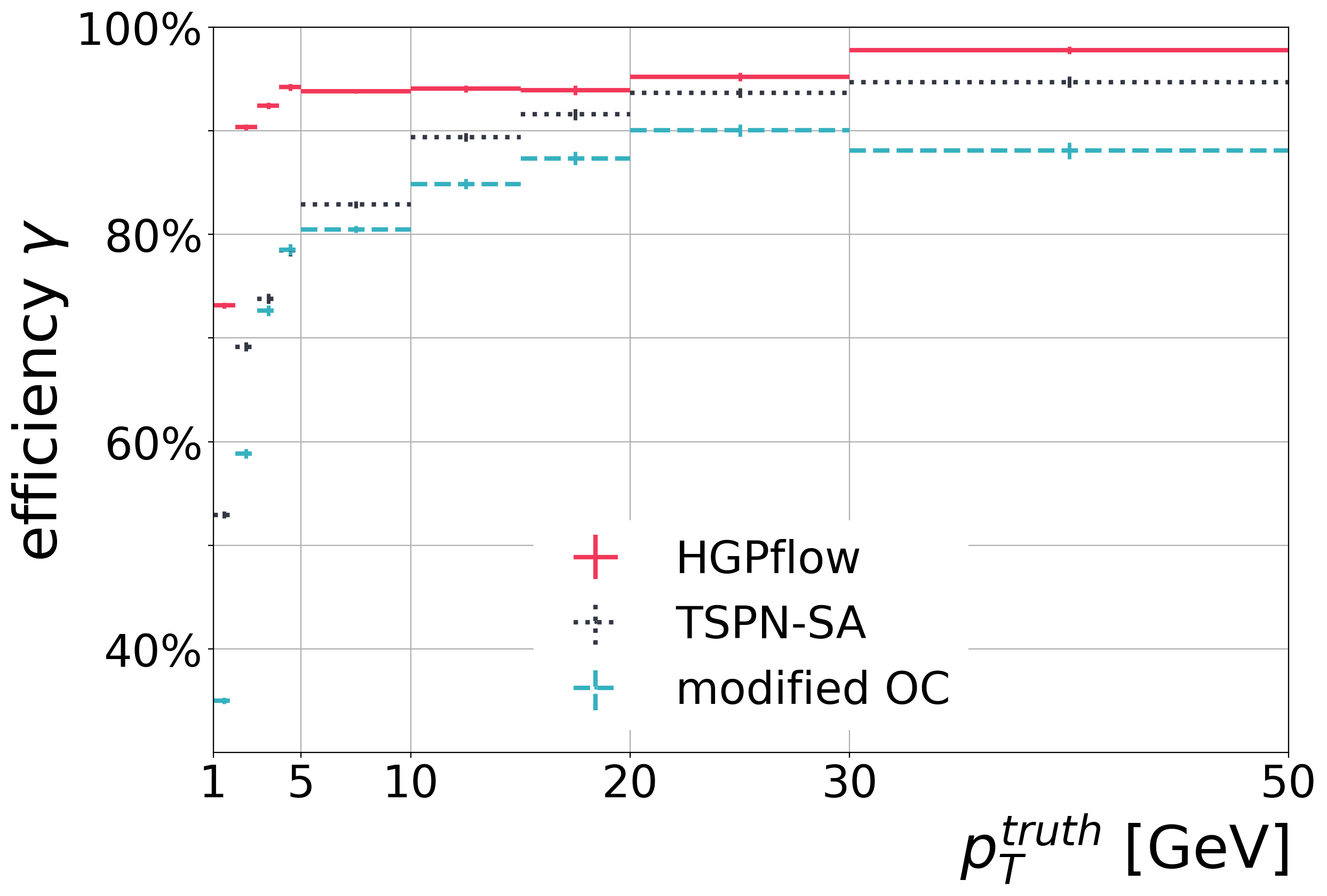}
    \caption{}
    \label{fig:eff_photon}
\end{subfigure}
\hspace{5mm}
\begin{subfigure}[b]{0.4\textwidth}
    \includegraphics[width=\textwidth]{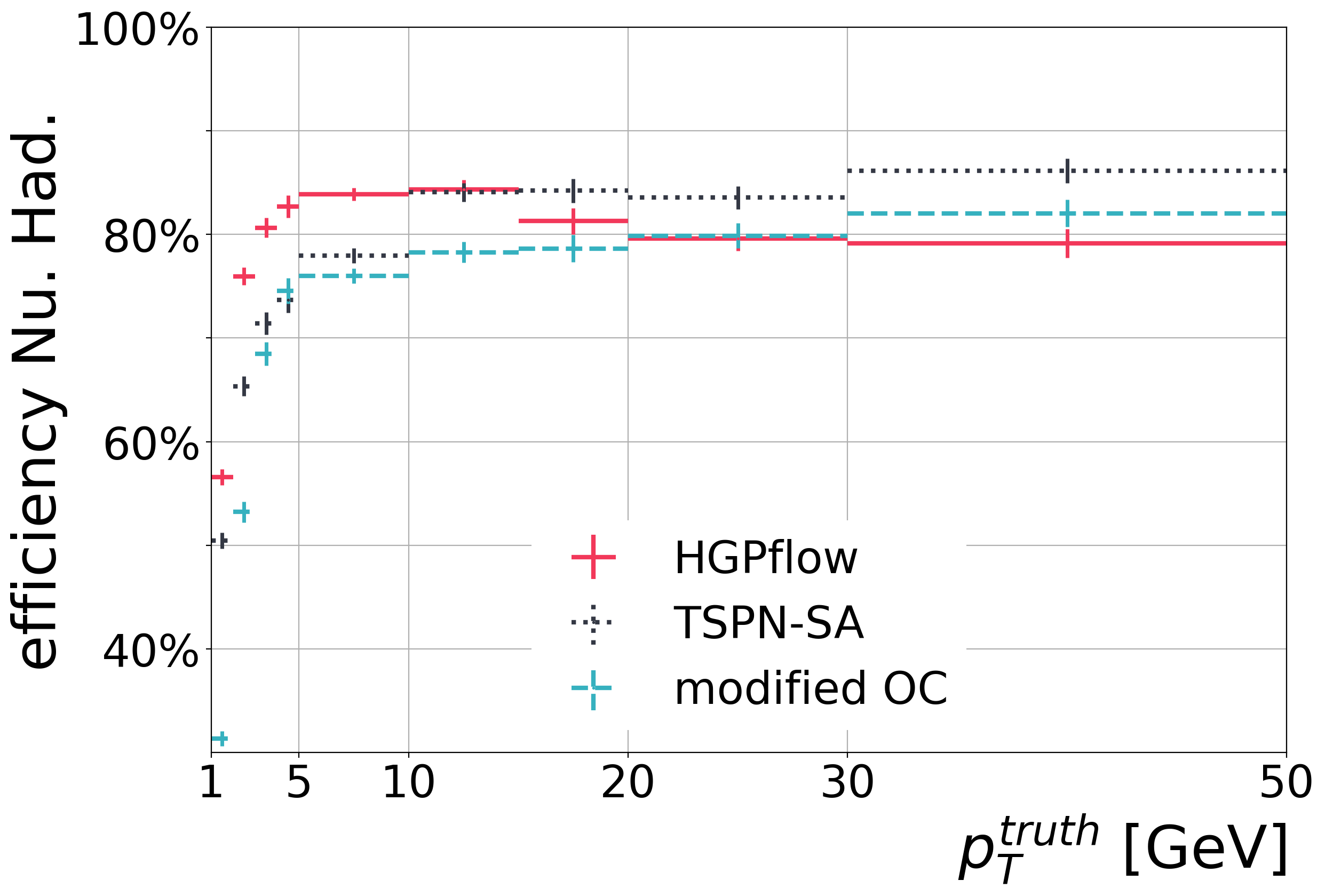}
    \caption{}
    \label{fig:eff_neutral}
\end{subfigure}
\begin{subfigure}[b]{0.4\textwidth}
    \includegraphics[width=\textwidth]{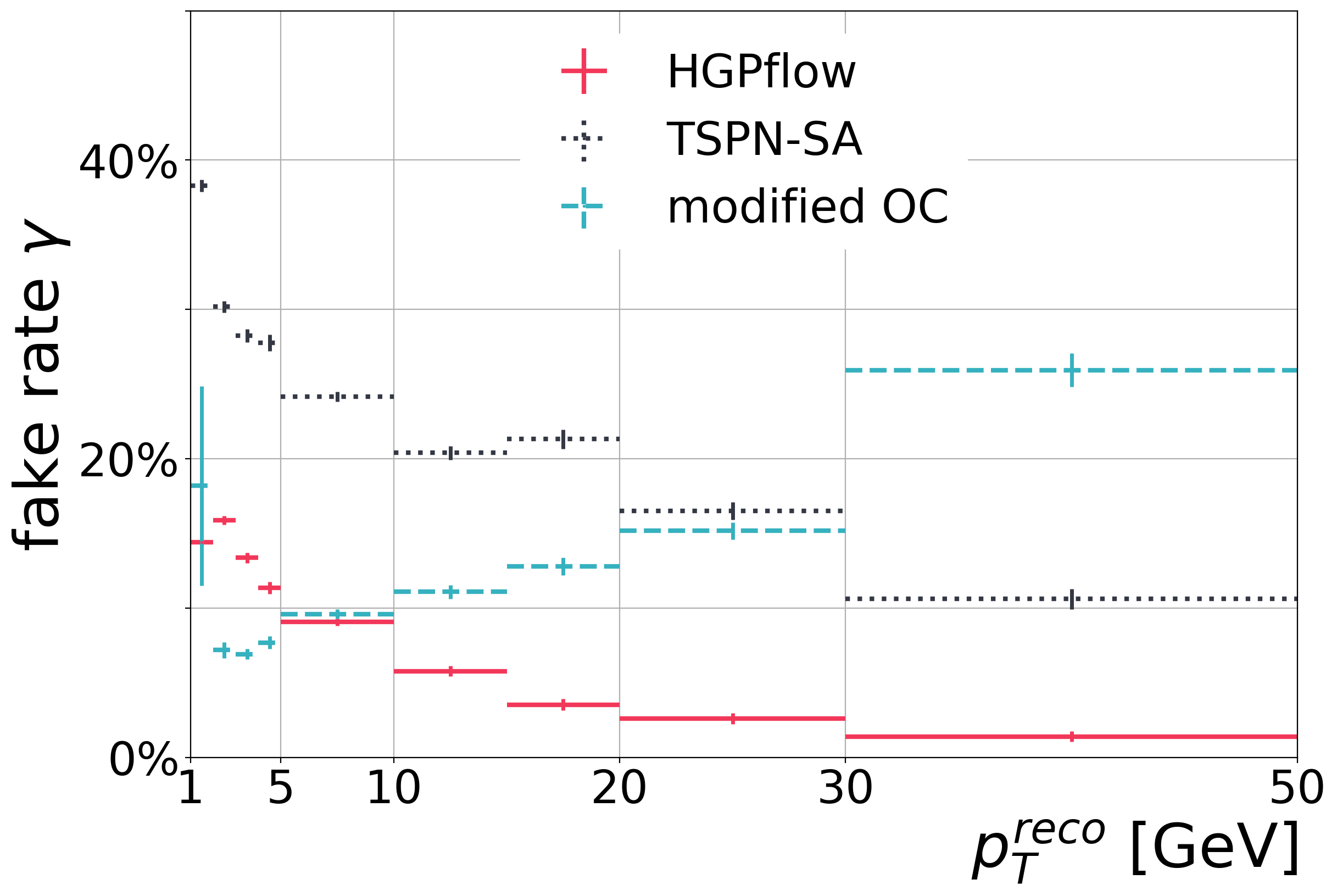}
    \caption{}
    \label{fig:fake_photon}
\end{subfigure}
\hspace{5mm}
\begin{subfigure}[b]{0.4\textwidth}
    \includegraphics[width=\textwidth]{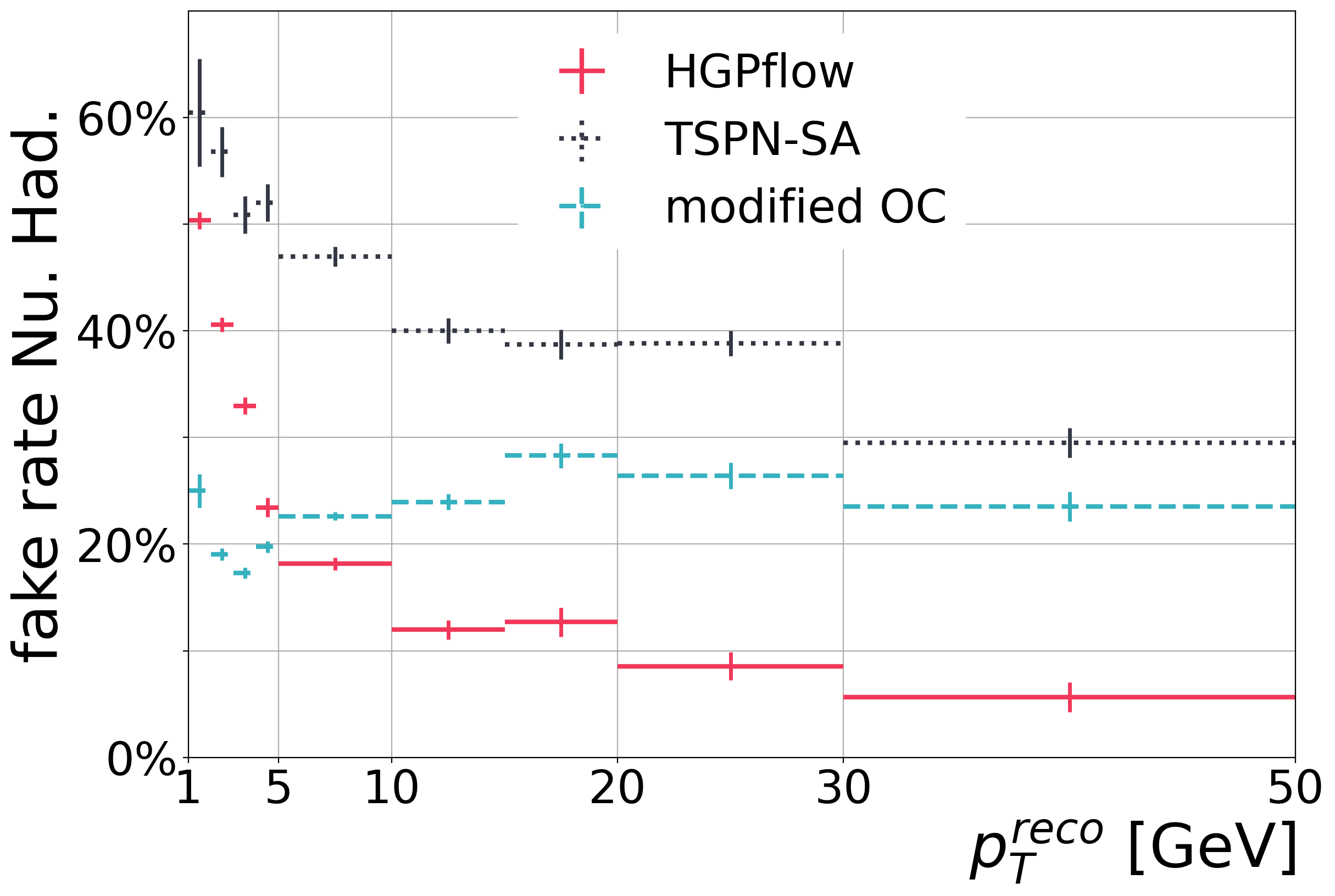}
    \caption{}
    \label{fig:fake_neutral}
\end{subfigure}
\centering
\begin{subfigure}[b]{0.4\textwidth}
    \includegraphics[width=\textwidth]{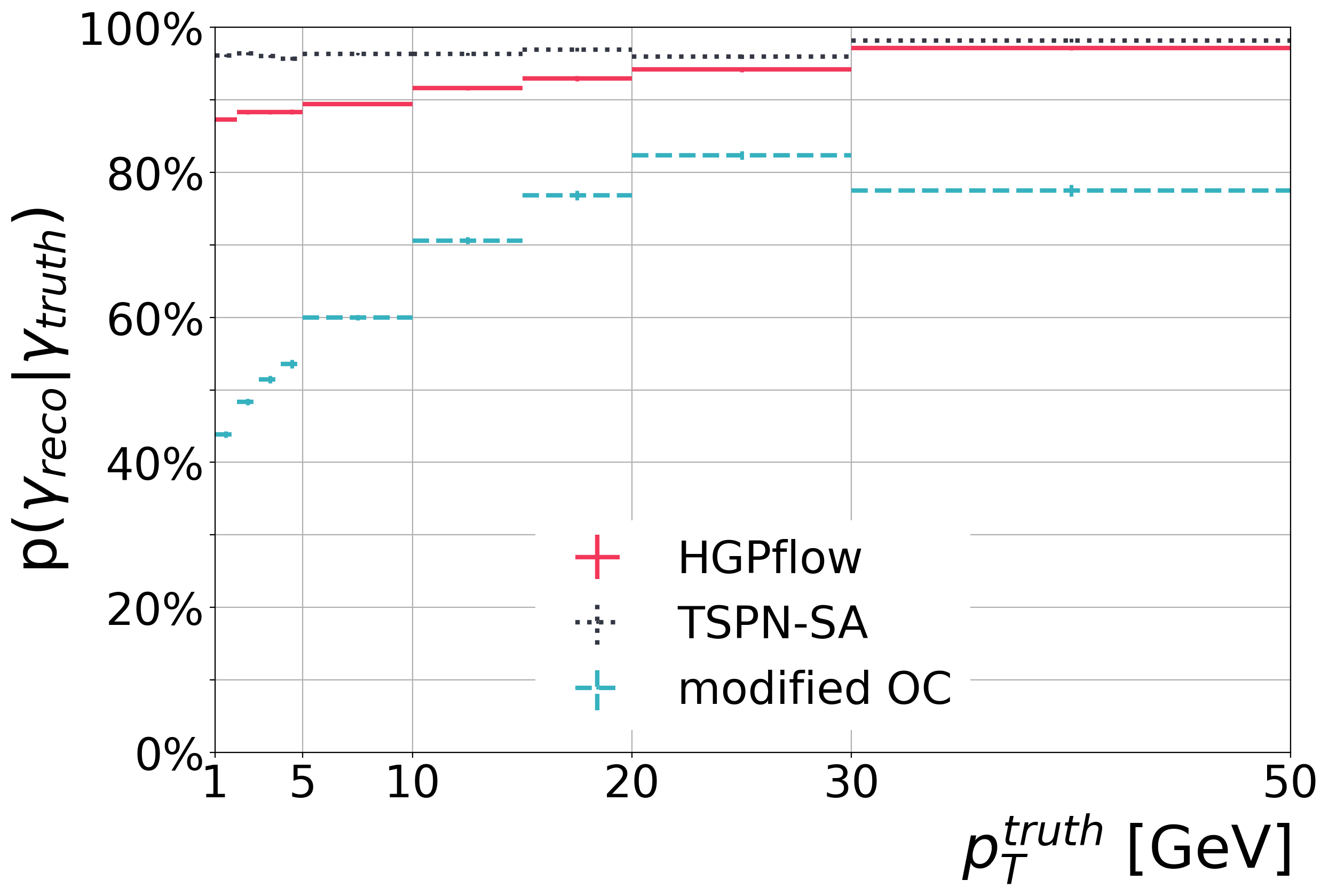}
    \caption{}
    \label{fig:p_photon}
\end{subfigure}
\hspace{5mm}
\begin{subfigure}[b]{0.4\textwidth}
    \includegraphics[width=\textwidth]{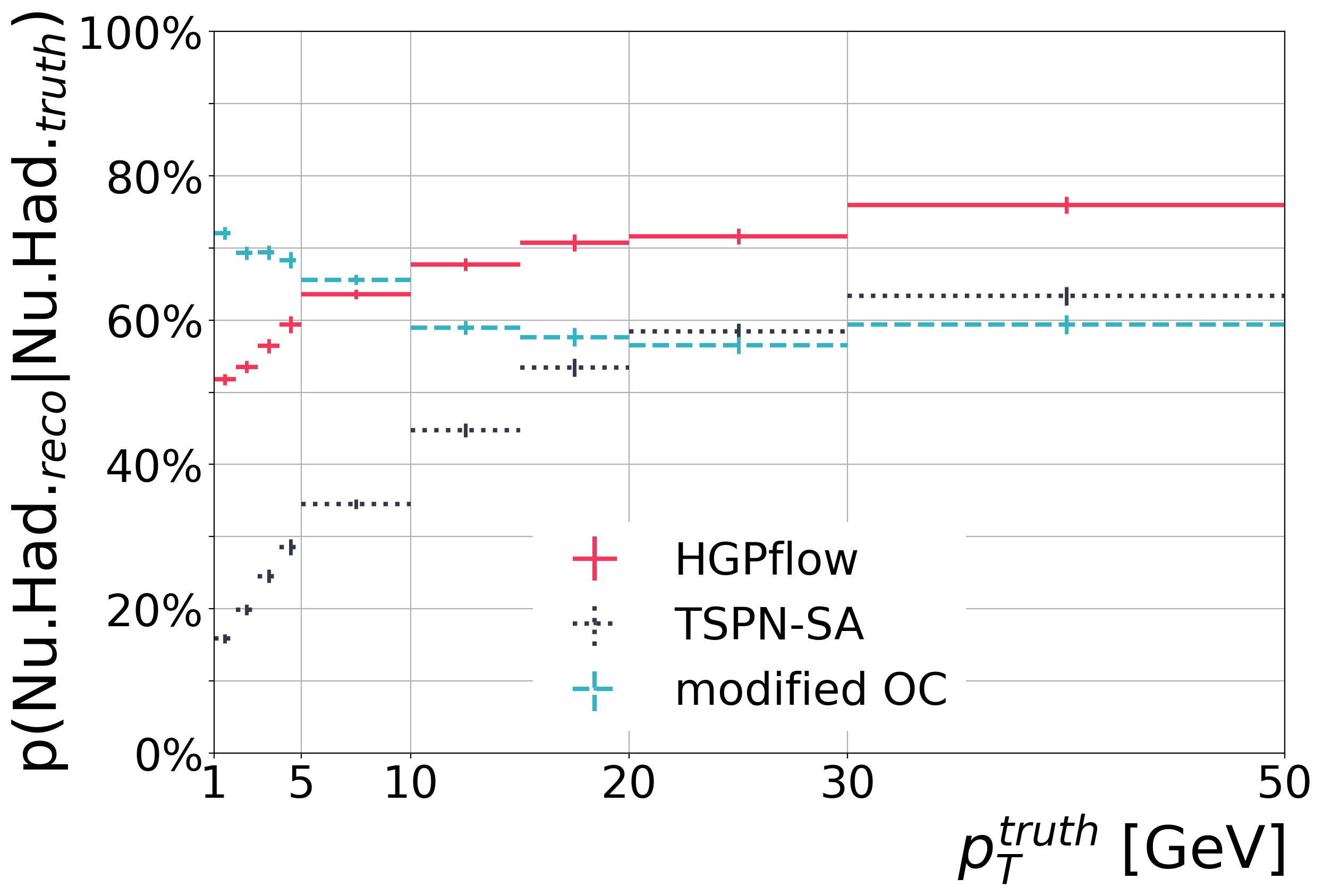}
    \caption{}
    \label{fig:p_neutral}
\end{subfigure}
\caption{Top: efficiency of matching predicted neutral particles to truth photons (a) and neutral hadrons (b) in a jet as a function of the associated truth particle $p_\mathrm{T}$. 
Middle: fake rate, i.e. probability that predicted photons (c) and neutral hadrons (d) are not matched to any truth neutral particle, as a function of the predicted particle $p_\mathrm{T}$.
Bottom: the probability that the predicted neutral particles which are matched to truth photons (e) and neutral hadrons (f) are assigned the correct class. For each curve, the misclassification probability is simply the difference of the curve from 1 .}

\label{fig:neutrals_eff_fr_p}
\end{figure*}

\begin{figure*}[ht!]
\centering
\begin{subfigure}[c]{0.45\textwidth}
    \includegraphics[width=\textwidth]{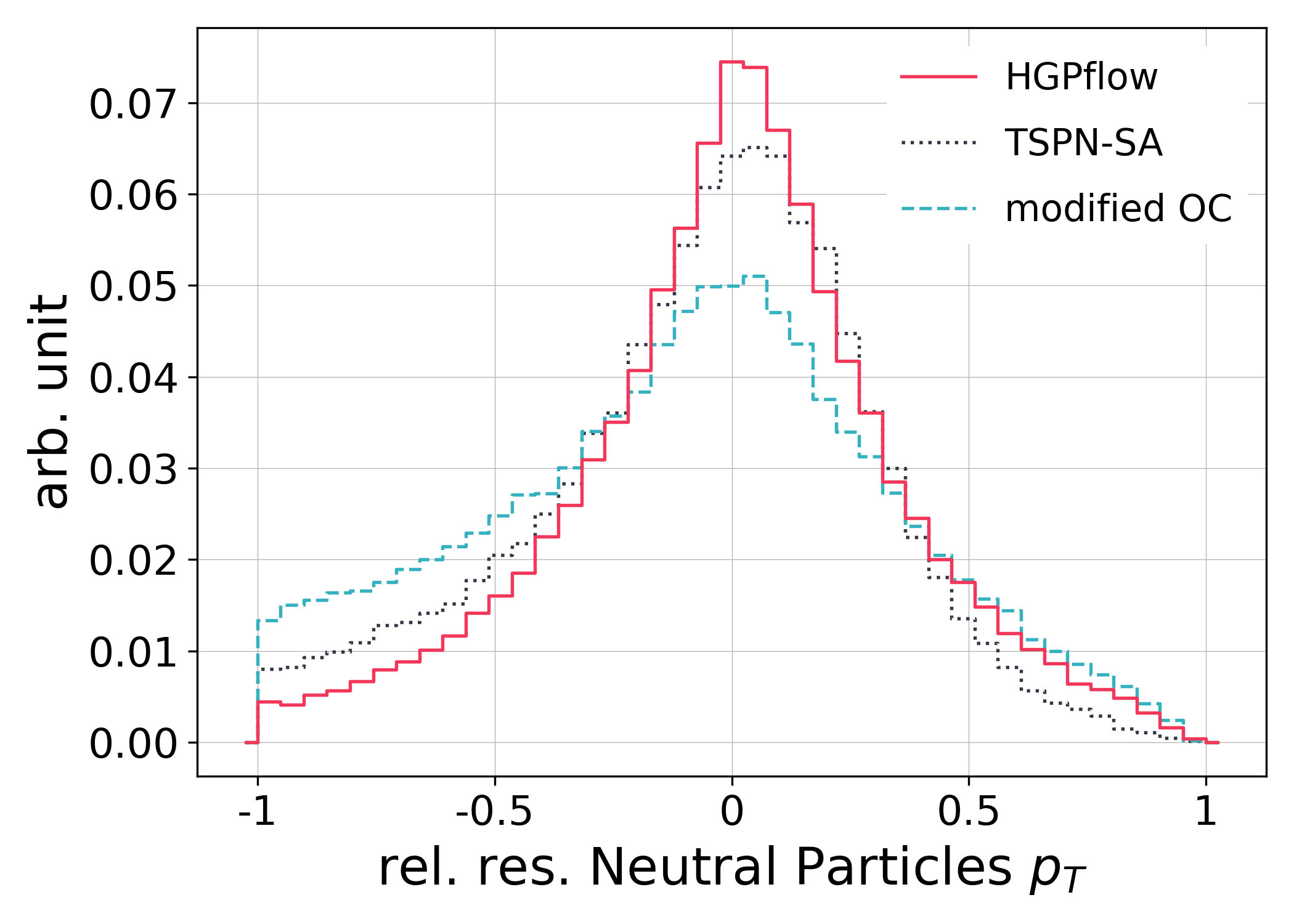}
    \caption{}
    \label{fig:neutral_residuals_pt}
\end{subfigure}
\\
\begin{subfigure}[c]{0.45\textwidth}
    \includegraphics[width=\textwidth]{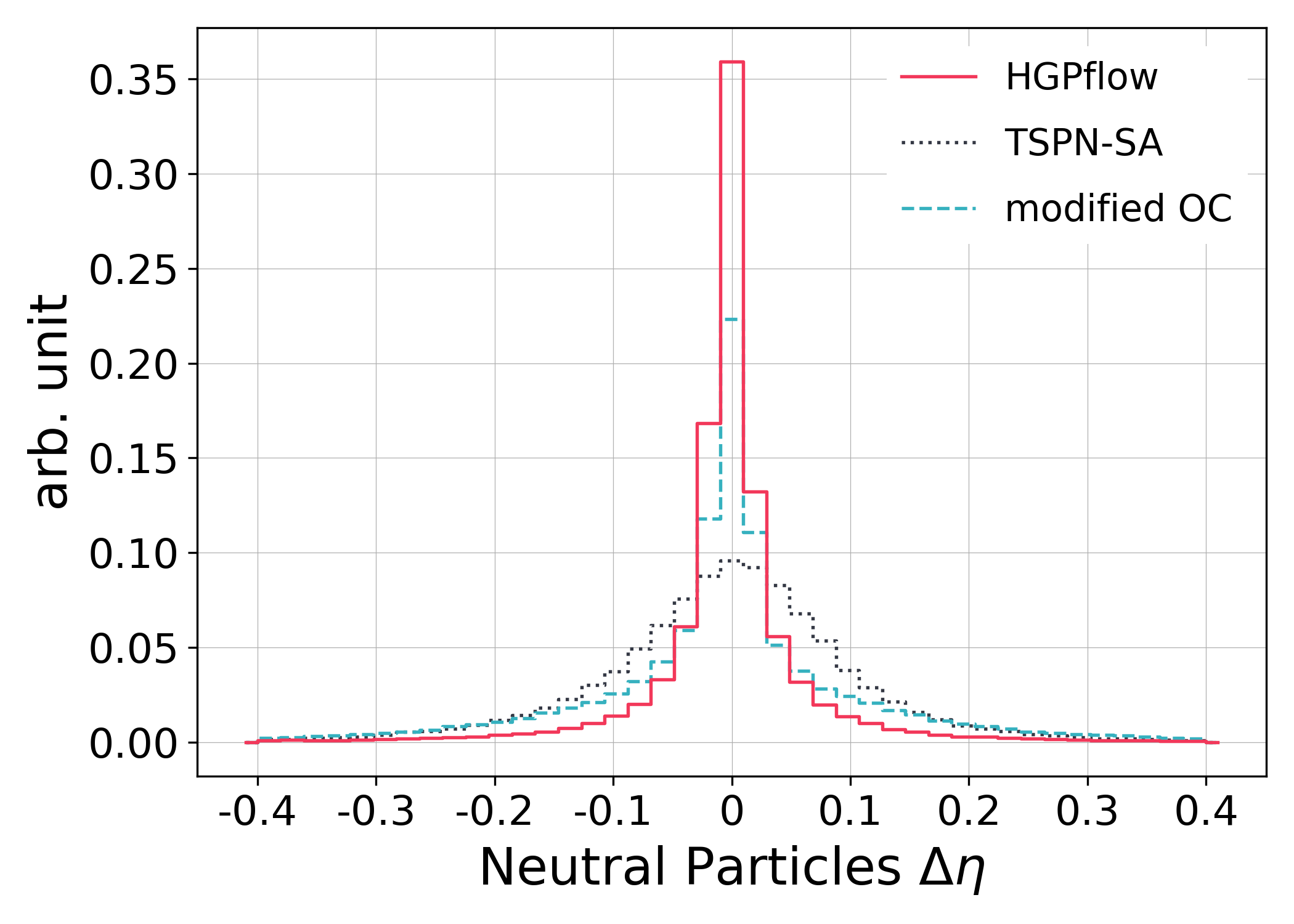}
    \caption{}
    \label{fig:neutral_residuals_eta}
\end{subfigure}
\begin{subfigure}[c]{0.45\textwidth}
    \includegraphics[width=\textwidth]{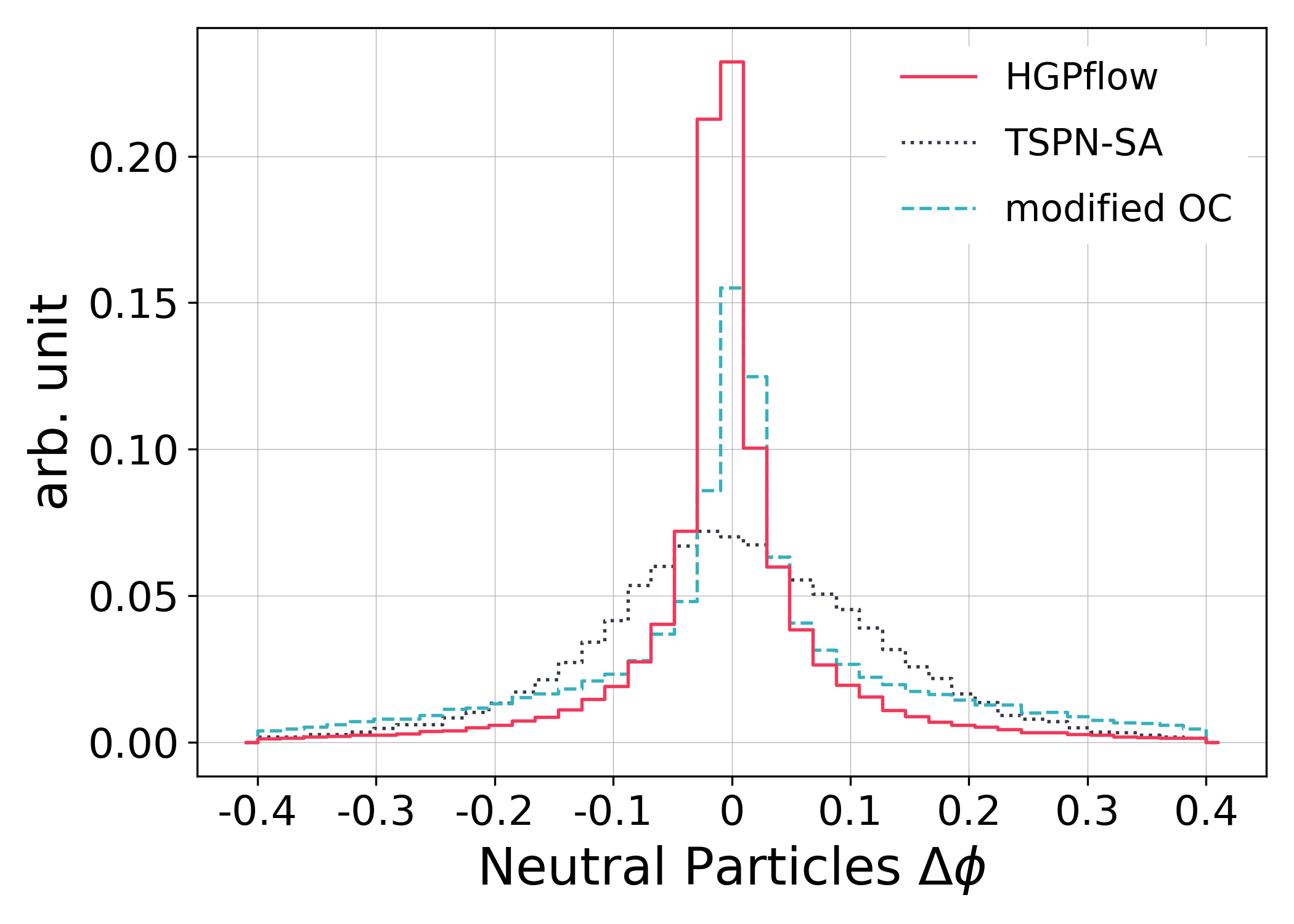}
    \caption{}
    \label{fig:neutral_residuals_phi}
\end{subfigure}
\caption{Distributions of relative residuals between predicted and true $p_\mathrm{T}$ (a), $\eta$ (b), and $\phi$ (c) of reconstructed neutral particles.}
\label{fig:neutral_residuals}
\end{figure*}


\subsection{Photon and neutral hadron performance}

The presence of photons inside jets is mainly due to decays of neutral pions and to a lesser extent from bremsstrahlung processes. Long-lived neutral hadrons on the other hand trace back to the shower of the initial partons. Disentangling these two components is not a trivial task and is detector-dependent -- for COCOA it is observed that 70\% of neutral hadrons below 5 GeV release all their energy in the ECAL layers, making it difficult to distinguish them from photons. This fraction steeply decreases with the energy of the initial hadrons to approach percent levels at around 20~GeV. Neutral particle reconstruction is further complicated inside the collimated environment because of the frequent overlaps between neutral showers. For this reason, efficiency and fake rate plots are computed by considering photon and neutral hadron predictions inclusively, without requiring a match between predicted and target class. 

Efficiencies, fake rates, and class prediction accuracy for photons and neutral hadrons are shown in Fig.~\ref{fig:neutrals_eff_fr_p}. The efficiency of reconstructing photons with $p_\mathrm{T}>2$ GeV is above 90\% for HGPflow, rising to 98\% for photons above $p_\mathrm{T}>30$ GeV. For \tspn{} and \oc{}, the efficiency for photons reaches 95\% and 90\%, respectively. Neutral hadrons above $5$ GeV are reconstructed with efficiencies ranging from 76\% to 86\% for the three algorithms. Fig.~\ref{fig:fake_photon} shows that the rate of producing unmatched photon predictions drops from 16\% (30\%) at a predicted $p_\mathrm{T}$ of 2 GeV to 1.4\% (11\%) above 30 GeV for HGPflow (\tspn{}). For neutral hadrons, the fake rate (Fig.~\ref{fig:fake_neutral}) is a factor of 2-4 times larger across the full $p_\mathrm{T}$ range for HGPflow, and 1.6-2.8 times larger for \tspn{}.

For the \tspn{} and HGPflow algorithms, reconstructing neutral particles at low-$p_\mathrm{T}$ is challenging because a large fraction of the target particles does not contribute a leading amount of energy to any topocluster in the event (33\% of photons and 25\% of neutral hadrons). In HGPflow, this is compensated in a supervised manner by learning subdominant contributions to topoclusters as fractional entries in the predicted incidence matrix. This limitation could be further overcome by using cell-level input nodes such as for the \oc{} algorithm.

Efficiency and fake rate plots are complemented by studying the probability of misclassification between photons and  neutral hadrons. In Fig. \ref{fig:p_photon} and Fig. \ref{fig:p_neutral}, both HGPflow and \tspn{} algorithms exhibit high accuracy of classification for predictions matched to photons, and for neutral hadrons an accuracy that rises with $p_\mathrm{T}$: from 51\% to 76\% for HGPflow and from 16\% to 63\% for \tspn{}. A lower accuracy is expected when considering that due to the class imbalance between photons and neutral hadrons of $5.8:1$ (inclusive), a random classifier would have an accuracy of roughly 15\% for neutral hadrons. Moreover, the class imbalance is $p_\mathrm{T}$-dependent, with the proportion of photons dropping off faster than neutral hadrons.

The \oc{} algorithm behaves similarly to the others for reconstruction efficiency of neutral particles, albeit with lower performance in most bins. For the fake rate, and in particular, for photons, \oc{} exhibits an increase with $p_\mathrm{T}$. The classification accuracy for \oc{} is also lower for photons and shows a different trend in $p_\mathrm{T}$ for neutral hadrons compared to HGPflow and \tspn{}. These differences are related to the fact that in \oc{} neutral particle predictions correspond to a subset of calorimeter cells passing the selection defined by the $t_b$ and $t_d$ threshold cuts. This introduces a sensitivity of the neutral particle cardinality to the number of cells per particle, which grows as a function of particle $p_\mathrm{T}$. The trend in fake rate appears to reflect this. Furthermore, towards high $p_\mathrm{T}$ a growing majority of cells belong to showers of charged particles, which makes the classification task more challenging in the \oc{} approach (which involves 3 classes, unlike HGPflow and \tspn{}). Introducing $p_\mathrm{T}$-dependent weights on the $\beta$ and $x$ prediction tasks could help counter this effect.



\begin{figure*}[ht!]
\centering
\begin{subfigure}[c]{0.45\textwidth}
    \includegraphics[width=\textwidth]{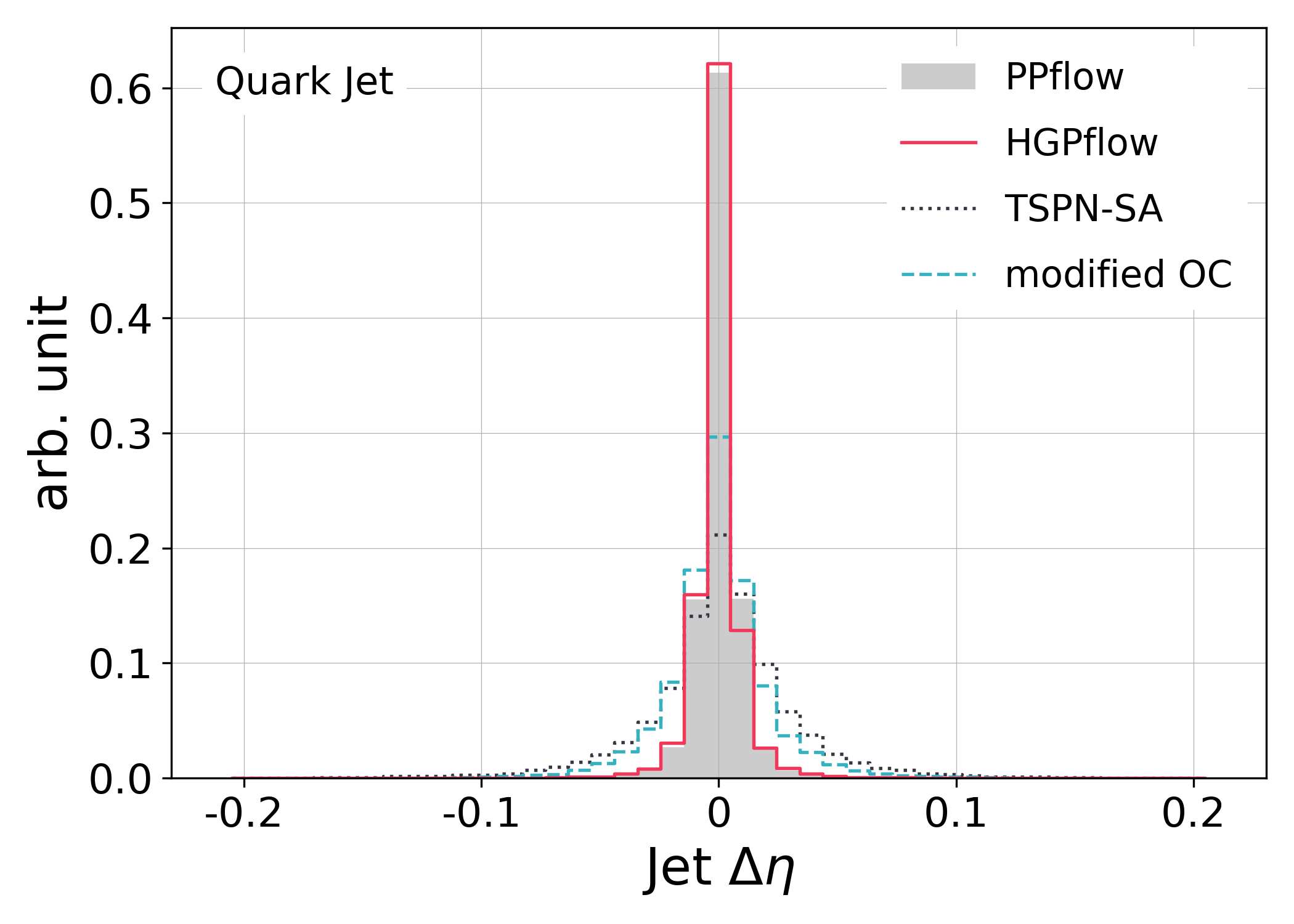}
    \caption{}
    \label{fig:jets_n1}
\end{subfigure}
\begin{subfigure}[c]{0.45\textwidth}
    \includegraphics[width=\textwidth]{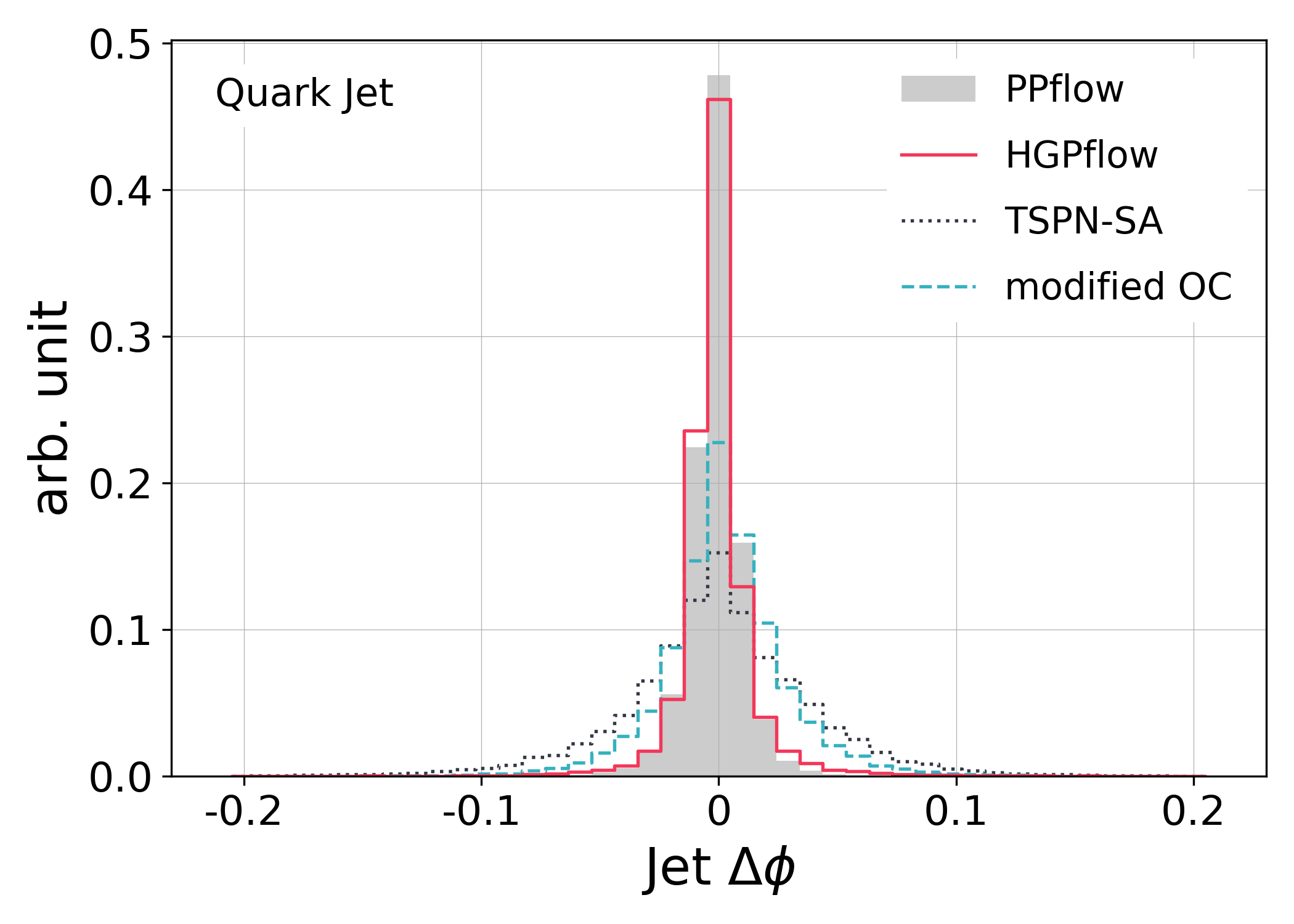}
    \caption{}
    \label{fig:jets_n2}
\end{subfigure}
\begin{subfigure}[c]{0.45\textwidth}
    \includegraphics[width=\textwidth]{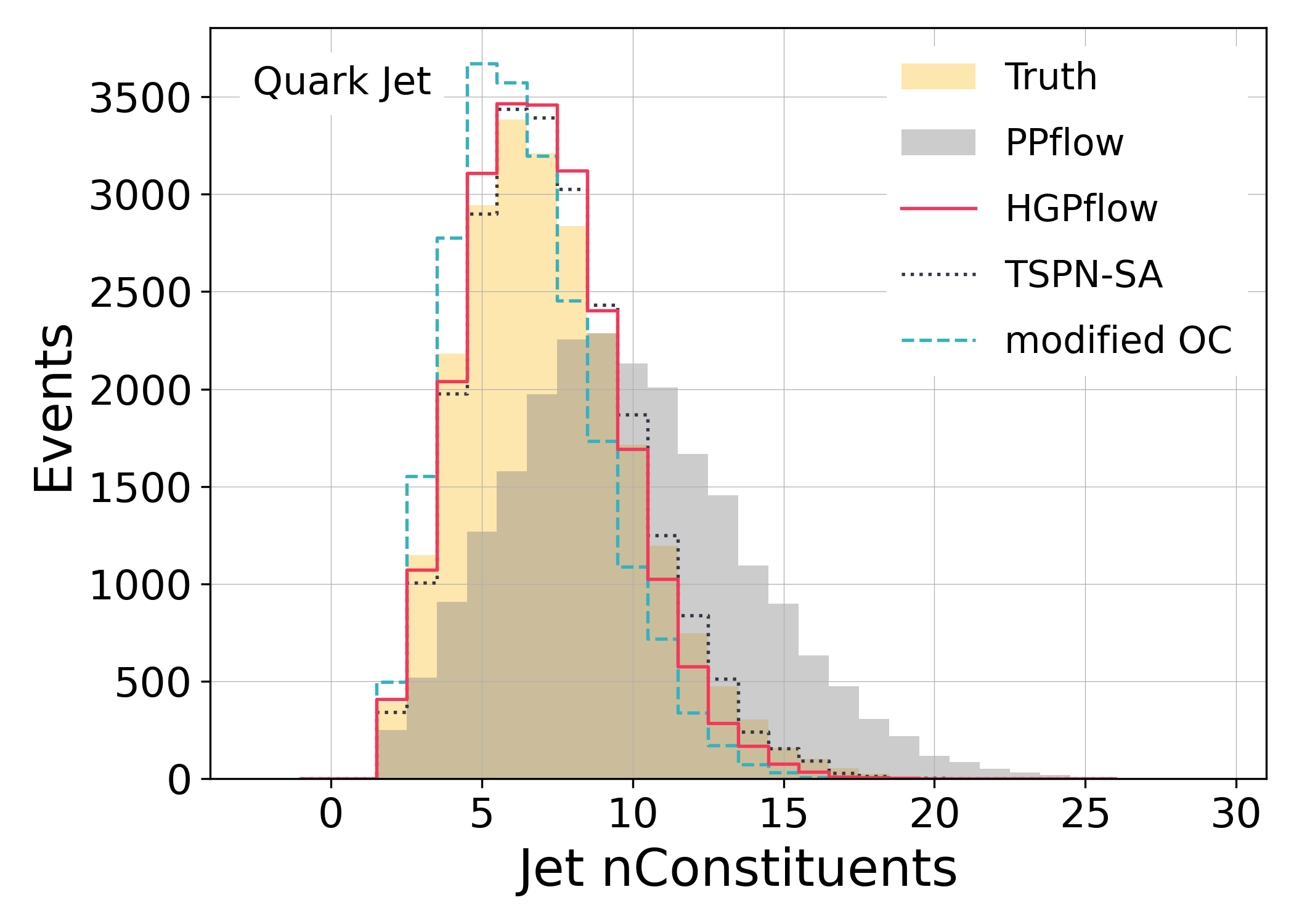}
    \caption{}
    \label{fig:jets_n}
\end{subfigure}
\begin{subfigure}[c]{0.45\textwidth}
    \includegraphics[width=\textwidth]{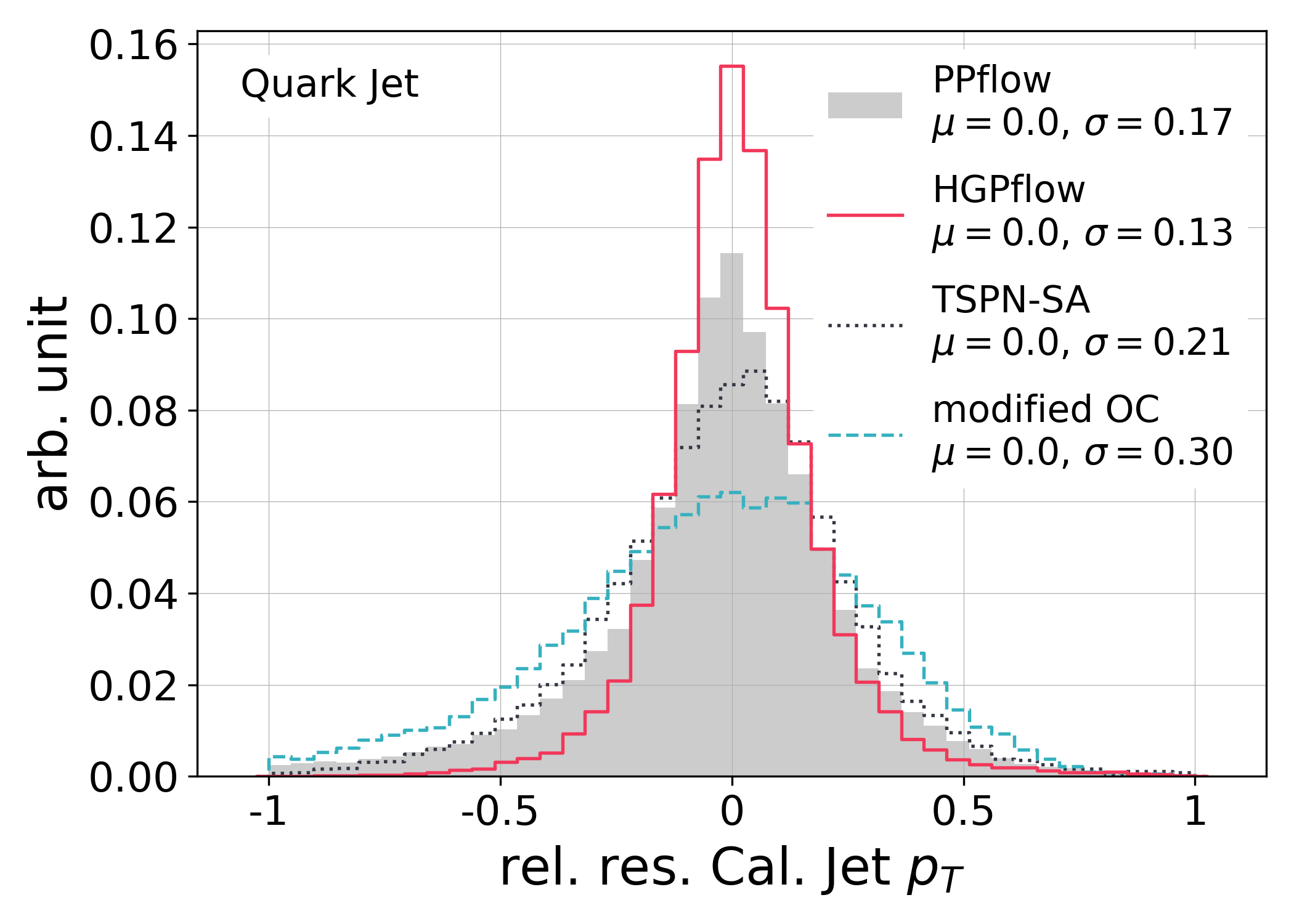}
    \caption{}
    \label{fig:jets_res_pt}
\end{subfigure}
\caption{Jet-level performance metrics shown for the three algorithms and the PPflow reference in comparison to true jets. Angular residuals between reconstructed and true jets (a and b), the number of jet-constituents in (c), and the calibrated $p_\mathrm{T}$ relative residuals in (d).}
\label{fig:jets_res}
\end{figure*}

Finally, relative residuals are used to quantify the ability to correctly predict the $p_\mathrm{T}$, $\eta$, and $\phi$ of the reconstructed neutral particles, shown in Figs. ~\ref{fig:neutral_residuals_pt}, \ref{fig:neutral_residuals_eta}, and \ref{fig:neutral_residuals_phi}, respectively.
The HGPflow algorithm shows the best performance at estimating accurately both angular variables and momentum. 
It is interesting to note that the \tspn{} algorithm has the worst performance for the angular variables. This is related to the usage of topoclusters in a less supervised way compared to HGPflow, where it is known which topoclusters contributed to the formation of a particle. The \oc{} algorithm instead uses the more granular calorimeter cells directly showing similar performance to the HGPflow algorithm for angular regression.

\begin{figure*}
    \centering
    \includegraphics[width=.90\linewidth]{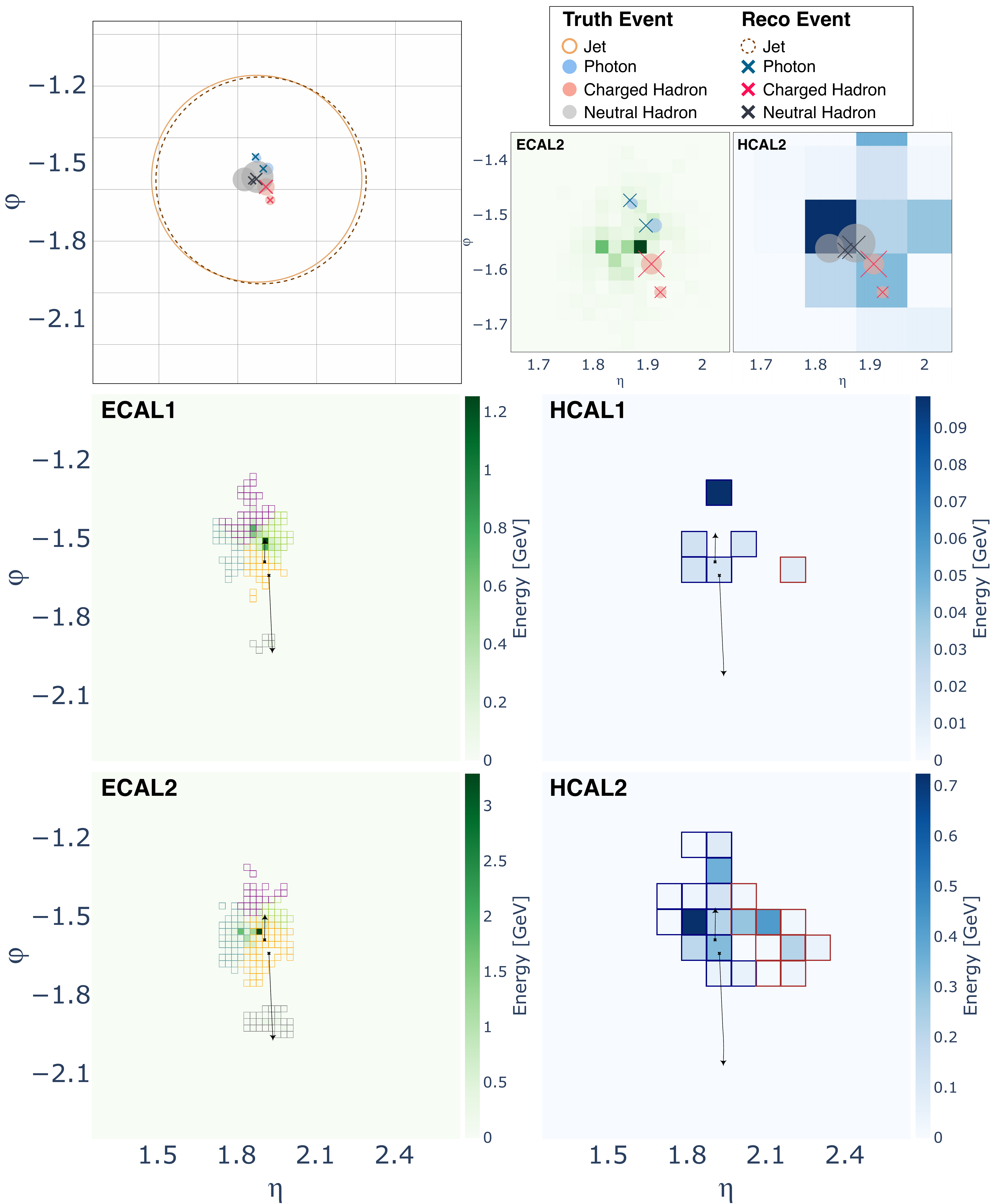}

    \caption{Event display of a single jet event from the test dataset. In the top left panel, the $\eta-\phi$ coordinates of truth particles with momentum above 1 GeV are shown as circles along with the set of predicted particles from the HGPflow algorithm shown as crosses. The set of particles and their $p_\mathrm{T}$ at truth level are as follows: two photons in blue (1.8, 3.0 GeV), a pair of neutral $K^0_L$ mesons in grey (12.3, 22.3 GeV), and two charged pions in red (2.2, 6.5 GeV). The circles of $R=0.4$ represent anti-$\mathrm{k}_t$ jets built from the truth (solid) and predicted (dashed) particle sets that nearly overlap in $\eta$-$\phi$ and have $p_\mathrm{T}$ agreement within $35\%$. In the top right panels, the truth and predicted particles are shown overlaid on a zoomed region of the ECAL2 and HCAL2 layers. In the bottom panels, the detector-level information serving as input to the reconstruction algorithms is shown for each of the first two layers of both ECAL and HCAL in the same $\eta-\phi$ plane. Cells that have the same border color belong to the same topocluster. Green and blue fill is used to indicate the energy of cells in the ECAL and HCAL layers, respectively. The arrows describe the tracks for charged particles from the interaction point with the arrowheads indicating the angular coordinate extrapolated at the given layer.}
    \label{fig:EventDisplay}
\end{figure*}

The models also perform differently for neutral particle $p_\mathrm{T}$ regression, shown in Fig. \ref{fig:neutral_residuals_pt}. The \oc{} model has a tendency to overestimate the neutral particle $p_\mathrm{T}$. A similar trend is less pronounced in the \tspn{} regression, while the distributions of predictions from HGPflow exhibit the least skew, in addition to the smallest mean and variance of the three. 

\subsection{Jet-level performance}
The ability to efficiently reconstruct jets and correctly predict their properties is a priority for experiments at the LHC. Jet performance depends on the overall efficiency, fake-rates and kinematic regression of the constituent particles, therefore being an important test for the ML reconstruction algorithms.

Following evaluation of the networks, jets are built using the anti-$\mathrm{k}_t$ algorithm \cite{Cacciari:2008gp} with a radius parameter of 0.4 and a minimum number of 2 constituents. We define three sets of jets with differing input constituents: 
\begin{itemize}
\item \textbf{Truth jets}: jets built using the set of target particles
\item \textbf{ML jets}: jets built using the sets of particles predicted by the OC, TSPN-SA, and HGPflow algorithms
\item \textbf{PPflow jets}:  jets built using tracks and topoclusters with the charged energy subtraction procedure of a parameterized particle-flow algorithm (see section \ref{sec:ppflow})
\end{itemize}
%

\begin{figure*}[ht!]
\centering
\begin{subfigure}[c]{0.45\textwidth}
    \includegraphics[width=\textwidth]{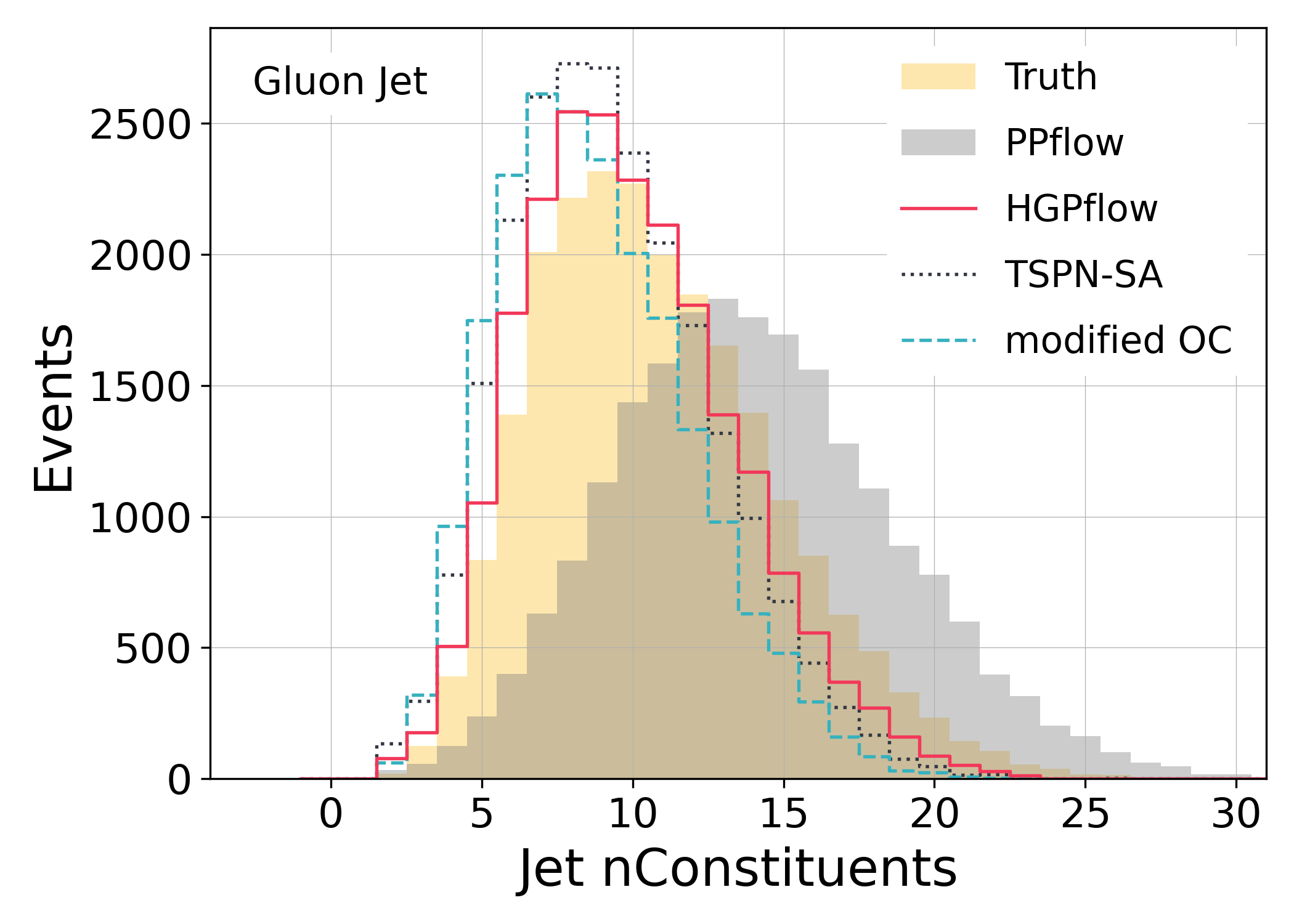}
    \caption{}
    \label{fig:gluon_n_constituents}
\end{subfigure}
\begin{subfigure}[c]{0.45\textwidth}
    \includegraphics[width=\textwidth]{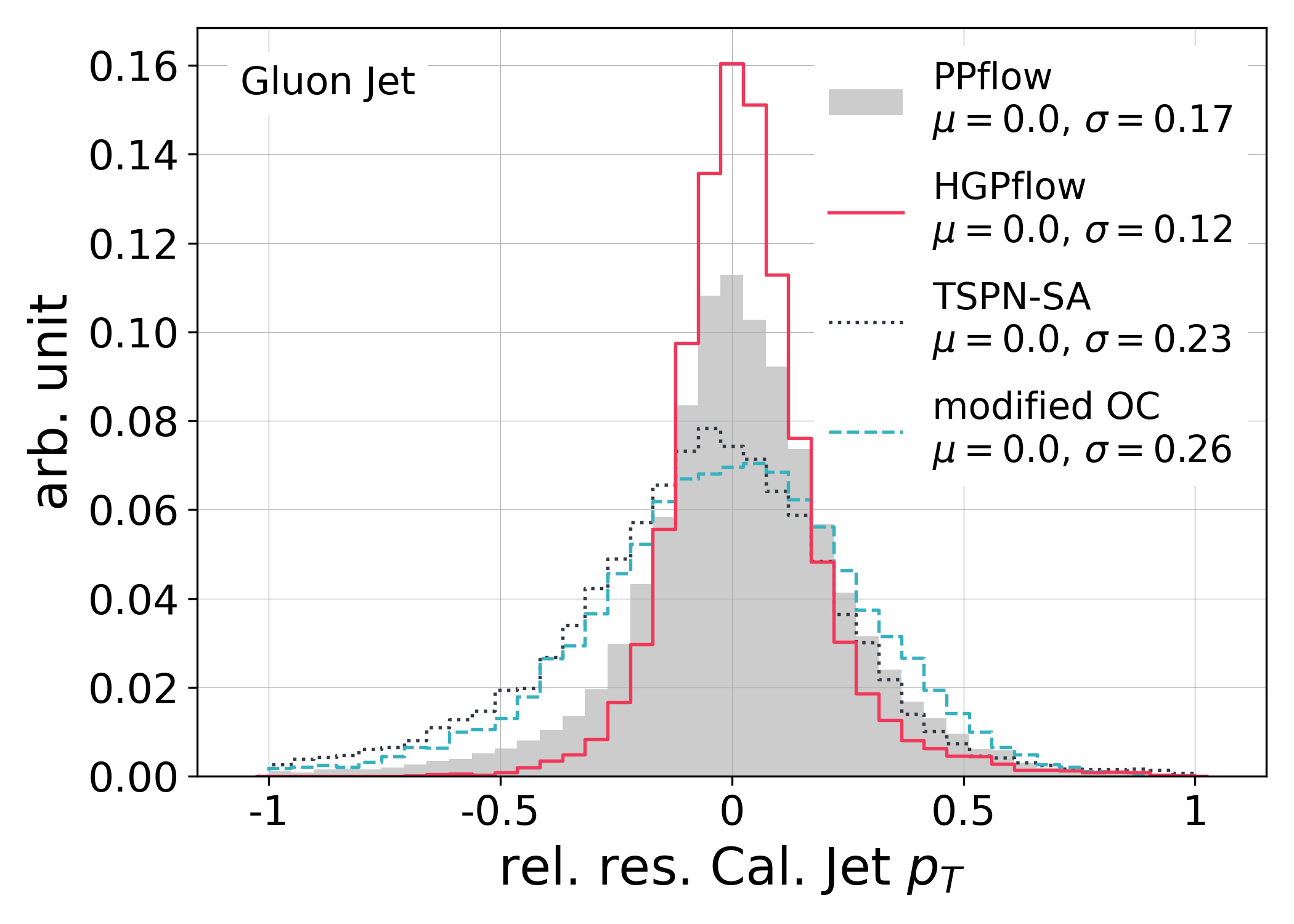}
    \caption{}
    \label{fig:gluon_pt_residual}
\end{subfigure}
\caption{Distributions of the number of constituents per jet (a), and distributions of relative residuals between predicted and true $p_\mathrm{T}$ (b) for the three ML particle reconstruction algorithms evaluated on the sample of gluon jets.}
\label{fig:gluon_residuals}
\end{figure*}

The number of constituents is shown in Fig.~\ref{fig:jets_n} for each algorithm and compared to that of the true jet. The ML algorithms, by accounting for neutral particles in the jet, are able to model this reasonably well. On the other hand, the PPflow distribution overestimates the truth distribution as expected, since its constituents are tracks and topoclusters rather than particles. 

In order to further optimize the jet-$p_\mathrm{T}$ and provide a more quantitative figure of the jet resolutions, a simplistic  calibration is applied. First jet $p_\mathrm{T}$ residual distributions are computed in different $p_\mathrm{T}$ bins. For each, a dedicated scale factor is computed. A functional fit is performed and the corresponding scale factor is applied to reconstructed jets based on their $p_\mathrm{T}$. This procedure is applied separately to each reconstruction algorithm.

Relative residuals are shown in Fig.~\ref{fig:jets_res}. As observed for neutral particles, HGPflow shows the best performance at the jet-level. In terms of jet angular observables, HGPflow is comparable to the traditional PPflow approach while for jet $p_\mathrm{T}$ resolution it shows a 24\% improvement relative to PPflow. The \tspn{} jet $p_\mathrm{T}$ resolution is better than \oc{}, while for angular observables \oc{} performs slightly better.

To help visualize the jet reconstruction task, Fig.~\ref{fig:EventDisplay} displays an event from the test dataset showing predictions from the trained HGPflow algorithm. In this example, each of the four neutral particles at truth level can be matched to a predicted particle with the correct class and an $\eta$-$\phi$ prediction consistent within the cell granularity. The calorimeter panels illustrate the tight arrangement of topoclusters used as input nodes for the HGPflow prediction.

\subsection{Performance on gluon jets}\label{sec:gluonperf}

To study the ability of the particle reconstruction algorithms to generalize beyond the training data to a new physics process, we evaluate the trained models on the dataset of single gluon jet events described in section \ref{sec:dataset}. The difference arising at the parton shower for gluon-initiated jets reflects itself in the dataset feature distributions, for example, the larger multiplicity of cells, tracks, and particles shown in Fig. \ref{fig:DataComposition}. Since the appropriate upper bound on the number of particles was determined based on the training dataset, a single gluon jet found to contain $>30$ particles was excluded from the dataset in order to evaluate HGPflow. The results for the three ML algorithms are shown along with the PPflow comparison in Fig. \ref{fig:gluon_residuals}. Overall, the algorithms demonstrate an ability to generalize: the number of predicted constituents is shifted slightly lower with respect to truth compared to the quark jet case (Fig. \ref{fig:jets_n}), while the jet relative $p_\mathrm{T}$ residual distributions are comparable to Fig. \ref{fig:jets_res_pt}. The rank of the algorithms in terms of performance remains the same as before, with HGPflow again boasting narrower jet $p_\mathrm{T}$ resolution than the PPflow comparison.

\section{Discussion}\label{sec:discussion}

\subsection{Perspective on the ML algorithms}\label{sec:algoimprovements}

Compared to the \oc{} and \tspn{} algorithms and the PPflow benchmark, HGPflow shows the best performance in terms of jet momentum resolution, which was not directly a training objective. This  traces back to superior modeling of neutral particle momentum shown in Fig.~\ref{fig:neutral_residuals_pt}. 
Unlike the other ML models, which must learn implicitly that a given energy deposit in the calorimeter cannot be associated with more than one parent particle, HGPflow benefits from being structured around the concept of energy conservation. Successfully predicting an incidence matrix defined via Eq.~\ref{eq:incidence} and the hyperedge indicator row entails knowing the energy contributions a given topocluster received from all particles (Fig.~\ref{fig:incidence}). Furthermore, the normalization ensures that energy attributed to a given particle candidate is not counted again in assignments to other particles. Since both the hyperedge representation and the proxy for neutral particle energy (Eq.~\ref{eq:proxy}) are weighted by entries of the incidence matrix, the property predictions which stem from these inputs inherit a bias towards energy conservation.


The hypergraph approach allows common elements of both the \oc{} and \tspn{} approaches to be handled in a more clear formalism. In the \oc{} potential loss (Eq.~\ref{eq:L_V}), a binary-valued incidence matrix $I_{ik}$ functions as a lookup table determining whether a node is repelled or attracted to the representation of a particle (i.e. condensation point). The clustering of nodes according to parent particle can thus be thought of as an indirect way of learning $I_{ik}$, limited by the extent to which the injective condition applies (discussed in section \ref{sec:models_common}). Likewise, the \tspn{} algorithm is built around an attention matrix between particle candidates $k$ and nodes $i$ from the input set which resembles an incidence matrix, although it is normalized along columns rather than rows. The attention weights also have a latent rather than physical meaning and are learned in an unsupervised way. On the other hand, HGPflow not only explicitly predicts the incidence matrix, which is the key to unraveling overlapping particle showers, but expresses it in the physical basis of energy contributions with the advantage mentioned previously.


We anticipate that the structure of HGPflow can be extended in at least three ways.
First, the input set granularity has been set without tuning to that of topoclusters from a standard ATLAS-like algorithm. This granularity can be increased to further enable the segmentation of overlapping energy deposits from nearby particles. Second, in the trainings for our results, the two objectives of incidence and properties prediction have been carried out nearly independently. However, a more powerful representation learning scheme could lead to a model which learns these two objectives in a synergistic way, allowing the incidence prediction to be informed by the properties prediction and vice versa. Finally, while Tab. \ref{tab:ModelComparison} indicates an acceptable inference time of HGPflow, more optimization is needed to reduce its training duration. This could be achieved by hyperparameter optimization in the recurrent training configuration and by exploring alternatives to the Hungarian matching, which the authors of \cite{zhang2021recurrently} identified as a computational bottleneck.

Similarly, besides the modifications proposed in \cite{qasim2021multi}, we suspect that the \oc{} algorithm can be substantially improved in future work. While the \tspn{} and HGPflow algorithms both have neural network layers for information exchange following the node encoding (i.e. successive attention and incidence weighted-updates), the node predictions of our \oc{} algorithm are likely limited by a comparatively narrow receptive field. One way to increase the receptive field in \oc{} is to add additional message passing blocks in the node encoder model, although a limited study of this option did not lead to conclusive improvement.
Another possibility is introducing an attention mechanism. The distance between nodes $i$ and $j$ in the latent clustering space entails a term $a_{ij} = x_i \cdot x_j$ with the form of attention, e.g. in the attractive potential, $\breve{V} \propto \Delta x^2 \ni -2 a_{ij}$. Therefore the clustering mechanism in \oc{} seems a natural point to introduce transformer blocks for enhanced information exchange, at the cost of computing additional gradients for the set of edges.

\subsection{Datasets for future work}\label{sec:newdatasets}

Several opportunities emerge for future investigation on new datasets. The performance reported in this work for $R=0.4$ quark and gluon jets suggests the application to substructure reconstruction in large-$R$ jets from boosted boson decays. The goal of studying particle reconstruction on a single jet dataset was to focus on the local system of overlapping particle showers which represents the kernel of the problem at the full-event level. For this reason, we envision that reconstructing full events could proceed by mapping the same trained model onto spatial partitions of the detector hits $D$ defined by topological and jet clustering algorithms, for instance. Given an effective scheme to deal with potential overlap, each partition could be treated as an approximately isolated set of input nodes, graph edges, and attention or incidence matrix weights. In this case the resource requirements reported in Tab. \ref{tab:ModelComparison} would scale linearly with the number of partitions. Ignoring overlap, an upper bound on the number of $R=0.4$ partitions of a full event could be estimated as $2\pi \cdot 6 / 0.4 \simeq 10^2$.

Studying the robustness of the models in the presence of pileup will also be an important follow-up task (see \cite{qasim2022end} for existing work in this direction). Likewise, the impact of interactions with material upstream of the calorimeter needs to be thoroughly addressed in a future dataset. In this case, electron pair production from early conversions and photons from bremsstrahlung will require a thoughtful definition of target particles to ensure that they can be feasibly distinguished during training. For example, photon conversions could be treated as a separate class, and electrons could be defined as targets depending on the quality of their associated track (if present at all).


\section{Conclusions}

In this paper, we applied ML-based reconstruction algorithms to the dense environment of a jet. Single-jet datasets were generated using a realistic calorimeter model to mimic the complexity and input features of data from the LHC. Particle-flow reconstruction is inherently a set-to-set task  suited for ML applications. Three ML algorithms -- a modified version of Object Condensation, a set-transformer architecture with slot attention (\tspn{}), and a novel hypergraph learning approach (HGPflow) -- were compared by their ability to reconstruct particles in the jet and jet-level quantities using an input graph comprising tracks and calorimeter clusters. In particular, the algorithms were trained to reconstruct individual neutral particles in the dense environment, a task going beyond the scope of traditional particle flow algorithms.

For charged particles, the algorithms learned to exploit the complementary information provided by calorimeter activity to improve on the measured track momentum. The efficiency of reconstructing photons and neutral hadrons reached 90\% and 80\%, respectively, for $p_\mathrm{T}>10$~GeV. The neutral particle fake rates were more variable for each algorithm, with the best performance being 10\% and lower for $p_\mathrm{T}>10$~GeV.

Jets formed from the predicted particles were compared to those from the true particles and also a parameterized particle flow baseline. HGPflow showed the best performance and surpassed the baseline in terms of both angular and momentum resolution of the jet. This can be explained from the fact that the hypergraph formalism is structured around energy conservation, which also makes its predictions more interpretable from a physics point of view. We anticipate that the suitability of the hypergraph formalism for the set-to-set task of particle reconstruction is yet to be fully leveraged. By demonstrating the potential of ML algorithms to disentangle the jet dense environment, our findings motivate the application to full collision events in future work.


\section{Acknowledgments}

The authors are grateful to Jan Kieseler for his advice on the \oc{} algorithm. ED is supported by the Zuckerman STEM Leadership Program. SG is partially supported by Institute of AI and Beyond for the University of Tokyo. EG is supported by the Israel Science Foundation (ISF), Grant No. 2871/19 Centers of Excellence. LH is supported by the Excellence Cluster ORIGINS, which is funded by the Deutsche Forschungsgemeinschaft (DFG, German Research Foundation) under Germany’s Excellence Strategy - EXC-2094-\\ 390783311.

\bibliography{pflow}
\bibliographystyle{unsrt}

\clearpage

\end{document}